\documentclass[aps,amsmath,amssymb,prb,twocolumn,superscriptaddress]{revtex4-1}

\usepackage{amsfonts}
\usepackage{amsmath}
\usepackage{graphicx}
\usepackage[usenames]{color}

\newcommand{\beq}{\begin{equation}}
\newcommand{\eeq}{\end{equation}}
\newcommand{\bsp}{\begin{split}}
\newcommand{\esp}{\end{split}}
\newcommand{\bpm}{\begin{pmatrix}}
\newcommand{\epm}{\end{pmatrix}}

\newcommand{\fig}[1]{Fig.~\ref{#1}}

\begin{document}

\title{Zeeman field induced topological phase transitions in  triplet superconductors}

\author{Timo Hyart}
\affiliation{Instituut-Lorentz, Universiteit Leiden, P.O. Box 9506, 2300 RA Leiden, The Netherlands}
\author{Anthony R. Wright}
\affiliation{School of Mathematics and Physics, University of Queensland, Brisbane, 4072 Queensland, Australia}
\author{Bernd Rosenow}
\affiliation{Institut f\"ur Theoretische Physik, Universit\"at Leipzig, D-04103, Leipzig, Germany}

%\date{September 21, 2012}

\begin{abstract}
We develop a general Ginzburg-Landau theory which describes the effect of a Zeeman field on   the superconducting order parameter in triplet superconductors. Starting from Ginzburg-Landau theories that describe fully gapped time-reversal symmetric triplet superconductors, we show that the Zeeman field has dramatic effects on the topological properties of the superconductors. In particular,  in the vicinity of a critical chemical potential  separating two topologically distinct phases, it is possible to induce a phase transition to a topologically nontrivial phase which supports chiral edge modes. Moreover, for specific directions of the Zeeman field, we obtain 
nodal superconducting phases with an emerging chiral symmetry, and with Majorana flat bands at the edge.  The Ginzburg-Landau theory is microscopically supported by a self-consistent mean-field theory of the doped Kitaev-Heisenberg model. 
\end{abstract}

%\pacs{73.43.-f,73.43.Cd, 73.43.Jn}

\maketitle

\section{Introduction}

%Topological stuff
Topological materials are characterised by the existence of protected gapless surface states and an insulating bulk. The importance of topological protection in condenced matter systems was originally realized in quantum Hall systems \cite{Thouless, Wen}
and superfluid helium \cite{Volovik-book}, but recently a number of other materials: topological insulators, superconductors and spin liquids, have become available for theoretical and experimental research \cite{HaKa10, Zhang-review, Kitaev06, Schny+08, Kitaev-class}. 

%Why are topological superconductors interesting
Topological superconductors are particularly appealing materials as they support non-Abelian quasiparticles: Majorana zero modes \cite{Lejinse12, Alicearev, Beenakker13, MooreRead, ReadGreen, Ivanov, KopninSalomaa91, Volovik, Wimmer, DasSarma06, Lee09, FuKane08, Sato09SO, Sau10, Kitaev-wire, FuKane09, Lutchyn10, Oreg, BuNa11, Vishwanath, Hyart12, Okamoto13, Scherer, Roy08,Qi2009, Sato09a, Sato09, Qi2010, Sato10, YTada, Chen14}. In addition to the fundamental interest arising from a new type of non-Abelian quantum statistics \cite{MooreRead, ReadGreen, Ivanov}, Majorana zero modes offer a promise for fault tolerant quantum computation \cite{Nayak+08, Alicea, Hassler11, Hyart13}:
Pairs of well separated Majoranas can be used to store quantum information non-locally (protecting the qubits from decoherence), the non-Abelian exchange statistics allows to perform certain quantum gates with exponentially small errors and the joint multi-qubit measurements can be implemented directly by measuring the total fermion parity of a selection of Majorana zero modes.

%Why to studyTRS invariant topological superconductors
In this Article, we concentrate on fully gapped triplet superconductors which are intrinsically two-dimensional time-reversal symmetric topological superconductors 
\cite{Hyart12, Okamoto13, Scherer, Roy08,Qi2009, Sato09a, Sato09, Qi2010, Sato10, YTada, Chen14}. We stress that this class of triplet superconductors is very important from the practical perspective, because they are favored by the weak-coupling $p$-wave pairing theory \cite{VollhardtWolfle}, and therefore one expects them to occupy significant portions of the phase-diagrams of the triplet superconductors. This expectation is strengthened by  recent self-consistent  mean-field calculations for doped Mott insulators \cite{Hyart12, Okamoto13}, which are described by  Kitaev-Heisenberg model \cite{JaKh09, Cha+10}. This type of two-dimensional helical $p$-wave superconducting phase can also appear in Sr$_2$RuO$_4$ interfaces \cite{YTada} or in BC$_3$ \cite{Chen14}. Moreover, these superconductors can be considered as two-dimensional analogs of the B phase of superfluid helium \cite{VollhardtWolfle}, so that our theory is  applicable to the planar phase of  $^3$He, which can become stable in thin films \cite{Silaev}. 

%The questions we want to address
Fully gapped  time-reversal invariant triplet superconductors are described by a vector order parameter 
\begin{equation}
\mathbf{d}(\mathbf{k})=\sum_i \eta_i \mathbf{d}_i(\mathbf{k}), \label{order_parameter}
\end{equation}
which is a linear combination of  basis vectors $\mathbf{d}_i(\mathbf{k})$. These basis vectors are degenerate solutions of the linearized gap equations \cite{Sigrist}, which describe the superconducting order parameter in the vicinity of the critical temperature. Therefore, there is necessarily a low-energy degree of freedom related to the coefficients $\eta_i$, which makes the order parameter $\mathbf{d}(\mathbf{k})$ more sensitive to external perturbations.  Because  $\mathbf{d}(\mathbf{k})$ describes the spin structure of the Cooper pairs, a particularly important perturbation is a Zeeman field. From a practical perspective, the Zeeman field can be  an externally applied magnetic field, an induced ferromagnetic order parameter, or  it can appear due to the coexistence of superconductivity and ferromagnetic order.  In the presence of ferromagnetic order,  the direction of the Zeeman field can at least partially be controlled with a weak external magnetic field. Possible materials where superconductivity and ferromagnetic order coexist have  attracted significant interest \cite{Mannhart, Moler, Pickett, Saxena}. In particular, there is a demand for a theory of the effect of a Zeeman field on superconducting order because of the recent exciting observations of coexistent ferromagnetic and superconducting order in two-dimensional LAO/STO interfaces \cite{Mannhart,Moler}. 

In this Article, we study self-consistently the effect of a Zeeman field $\mathbf{b}$ on triplet superconductors. We concentrate on a rather generic situation, where the basis vectors are perpendicular to each other
%******************************  structure of basis functions ******************
\begin{equation}
\mathbf{d}_i(\mathbf{k})\ = \ f_i(\mathbf{k}) \hat{\mathbf{e}}_i  
\end{equation}
%*****************************************************
%
and $f_i(\mathbf{k})$ are real functions, which describe the momentum dependence of a $p$-wave order parameter.
In this case, we can understand the effect of a Zeeman field on the relative energetics of the order parameters  by recalling that the electron spins  in the condensate are confined to a plane perpendicular to the $\mathbf{d}$-vector. Thus, a Zeeman field pointing in the direction  $\hat{\mathbf{e}}_i$ will try to confine the electron spins in a direction different from  that preferred by the order parameter  basis function $\mathbf{d}_i$, and therefore it will energetically disfavour this basis function relative to the other ones. This qualitative insight can be encoded in a contribution to the free energy  $\propto \big(\mathbf{b} \cdot \mathbf{d}(\mathbf{k})\big)\big(\mathbf{b} \cdot \mathbf{d}(\mathbf{k})^*\big)$, which needs to be integrated over the Brillouin zone with a weight function. Such a contribution to the free energy is also obtained by using the expression for the spin susceptibility of a triplet superconductor 
 \cite{VollhardtWolfle} to derive the contribution to the free energy  quadratic in the Zeeman field.  
Assuming that  the integrated basis functions $f_i(\mathbf{k}) f_j (\mathbf{k})$ satisfy a permutation symmetry,  we expect that the free energy of the superconductor contains an anisotropic mass term  of the form
%
%**************************   magnetic field induced mass term  ***********
\begin{eqnarray}
&&D_2^A \sum_i b_i^2 |\eta_i|^2 -D_2^B \sum_{i<j} b_i b_j (\eta_i \eta_j^*+\eta_j \eta_i^*) \ \ , \label{anisotmass}
\end{eqnarray}
%************************************************************************
%
where $D_2^A$ and $D_2^B$ are constants. 
 
Depending on the physical realization of the fully gapped time-reversal symmetric triplet superconductor, we distinguish two qualitatively different situations, which are characterized by the number of independent basis vectors and corresponding pairing wave functions. For the wave functions $f_i(\mathbf{k})$  we can define a scalar product 
$\langle f_i | f_j\rangle $ by integrating them around the Fermi surface  with a weight functions [see the last line in Eq.~(\ref{coeff})], and in this way define the concept of orthogonality between them. In the case of large doping, the Fermi surface can be approximated to be circular around the $\Gamma$-point, and a continuous rotational symmetry emerges from the discrete lattice  symmetry. Since p-wave pairing corresponds to a total angular momentum $L=1$, there are three basis wave functions with quantum numbers $l_z =-1, 0, 1$ available. Denoting the polar angle of the momentum $\mathbf{k}$ by $\varphi$,  for a two-dimensional superconductor only the two functions 
$e^{\pm i \varphi}$ with angluar momentum $l_z = \pm 1$ correspond to motion in the $x$-$y$-plane. 
Hence, there can be only  two orthogonal  pairing wave functions $f_i(\mathbf{k})$ for a two dimensional superconductor. On the other hand, for the case of three basis vectors $\mathbf{d}_i(\mathbf{k})$, the pairing wave functions cannot be all orthogonal to each other.

The case of two independent basis vectors is relatively easy to understand, and will be discussed first. As discussed above, the basis functions $f_i(\mathbf{k})$ can be orthogonal to each other, so that the coefficient $D_2^B=0$, and the effect of Zeeman field is just described by $D_2^A (b_1^2 |\eta_1|^2+b_2^2 |\eta_2|^2)$.  Thus,  the Zeeman field generically tries to make the superconductor anisotropic, while at the same time    the superconductor remains fully gapped for reasonably small Zeeman fields. The only exceptions are gap closings which may occur in the high-symmetry points of the Brillouin zone $\Gamma$ and $M_i$, where the order parameter always vanishes. The gap closings occurring in these points result in topological phase-transitions, where the Chern number describing the number of edge modes changes.  This type of phase transitions were discussed earlier in Ref.~\onlinecite{Sato09} without taking into account the self-consistent changes in the superconducting order parameter. We find that a possible anisotropy of the order parameter is not important for these transitions, and therefore it is possible tune the superconductor from a helical to a chiral phase also when the self-consistent changes are taken into account. 
 
On the other hand, when  the Zeeman energy  becomes comparable to the superconducting gap,  the nature of the superconducting phase depends on the strength and direction of the Zeeman field.   Recently, without considering self-consistent changes in the order parameter, it was proposed that a Zeeman field in the $(x,y)$-plane could induce a transition from a fully gapped helical $p$-wave superconductor to a gapless phase with Majorana flat bands \cite{Lee12}. The existence of this transition is tightly connected to the presence  of a chiral symmetry, which is always guaranteed in a time-reversal symmetric situation. 
However, with a time-reversal breaking 
 external  Zeeman field, a chiral symmetry  is only  present for specific relative directions of  order parameter vector and  Zeeman field. In the case of a two-component triplet superconductor, it is possible to show that a chiral symmetry exists for all Zeeman fields in the $(x,y)$-plane \cite{Lee12}.   The importance of the chiral symmetry is twofold here. First, the chiral symmetry guarantees that the superconductor becomes gapless when the magnitude of the Zeeman field is larger than the superconducting gap. Secondly, it allows the definition of a lower dimensional topological invariant, which describes the number of Majorana flat bands on the edge of the sample in a momentum space region corresponding to the interval in between the projections of the bulk nodal points on the edge. 
Such flat bands \cite{Heikkila-flat-bands} are known to exist in graphene \cite{Nakada96, Fujita96} and intrinsic nodal superconductors \cite{CuTheory, Ryu02, Volovik-vortices, Tanaka10, SchnyderRyu11, Brydon11, Sato11, Schnyder12}. 

We find that the topological phase transition from a fully gapped helical $p$-wave superconductor to a gapless phase with Majorana flat bands survives also when the self-consistent changes in the order parameter are taken into account. For generic Zeeman field directions in the $(x,y)$-plane, the flat bands are qualitatively similar to the  ones discussed in Ref.~\onlinecite{Lee12}, and within a weak coupling theory of superconductivity, the flat bands appear only in a tiny interval of momentum around the Fermi momentum $k_F$. On the other hand, a dramatic change occurs once the Zeeman field is applied along one of the symmetry axis, say along the $x$-axis, and when it is strong enough so that $\eta_1$ becomes zero. Then, an even number of flat bands appears over a wide range of momenta connecting $-k_F$ to $k_F$, and these flat bands are absent if the self-consistent changes in the order parameter are neglected. The difference with respect to the previous case  \cite{Lee12}  is the emergence of a new type of chiral symmetry. Furthermore, this second type of chiral symmetry guarantees the existence of  flat bands also when the Zeeman field is in the $(x,z)$-plane. Thus, the  self-consistent changes in the order parameter considerably widen the parameter space for observing the Majorana flat bands.

If one has three independent basis vectors in Eq.~(\ref{order_parameter}), the allowed parameter space for the order parameter is significantly expanded, resulting in a richer dependence of the order parameter on the Zeeman field. In particular, in 2D systems necessarily $D_2^B \ne 0$. Based on the expression (\ref{anisotmass}), we expect that a Zeeman field with three nonzero components $b_i$ of comparable magnitude will choose an optimal direction of the order parameter in $(\eta_1, \eta_2, \eta_3)$-space, so that all the components $\eta_i$ will remain nonzero also in the presence of Zeeman field,  until the Zeeman field reaches a critical value at which the superconductivity vanishes.  In this case, one might expect that the superconductor is generically fully gapped and this expectation is supported by  numerical evidence for the Kitaev-Heisenberg model discussed below. The only exceptions are the specific values of the Zeeman field where topological phase transitions take place due to a change in the Chern number. 
For the Kitaev-Heisenberg model, we find that this way it is possible to induce a topological phase-transition to a time-reversal broken superconducting phase that supports unpaired Majorana zero modes at the edges and within vortices, and that this phase exists for a wide range of directions and strength of the Zeeman field.

If only two components of the Zeeman field are nonzero, say $b_1$ and $b_3$, then $\eta_1$ and $\eta_3$ have an enhanced mass in the free energy, so that by increasing the Zeeman field  one can reach a situation where $\eta_1 = \eta_3 = 0$, and only $\eta_2$ is nonzero. In this case, the superconductor is necessarily nodal due to the $p$-wave nature of the basis functions, and it satisfies the same emergent chiral symmetry which is found in the two-component triplet superconductor. 
Going further, one can consider a situation in which 
only one component of the Zeeman field is nonzero, say $b_3$. Then, $\eta_3$ is suppressed relative to the other two components, and one can reach a situation where $\eta_3=0$ and the other two components, $\eta_1$ and $\eta_2$, are nonzero.  Furthermore, equation (\ref{anisotmass}) does not specify the phase difference between $\eta_1$ and $\eta_2$ in this situation. However, because the system tries to align the Cooper pair spin along the axis of the Zeeman field, the free energy also contains a term $\propto i \xi(\mathbf{k}) \mathbf{b} \cdot \mathbf{d}(\mathbf{k}) \times \mathbf{d}(\mathbf{k})^*$, which needs to be integrated over the Brillouin zone with a suitable weight function. This term is typically very small, because the contributions with $\pm \xi$  approximately cancel each other around the Fermi surface, but for sufficiently strong Zeeman fields this term still locks the phase difference between $\eta_1$ and $\eta_2$ in such a way that a nodal phase with a chiral symmetry appears. Putting everything together, we can discuss all  in-plane directions in a common framework, in which  however the critical field $b_{c, \rm{ch}}$ for the appearance of the nodal superconducting phase with chiral symmetry depends on the direction of the Zeeman field. Similarlarly to our previous discussion, we find  Majorana flat bands at the edge of the system, but we stress that here they always originate from self-consistent changes in the superconducting order parameter, and they are absent if these changes are not taken into account. Thus their physical origin is very different from the flat bands discussed in Ref.~\onlinecite{Lee12}.

%Structure
The structure of the paper is as follows.
In section II, we describe the general Ginzburg-Landau free energy expansion of a spin-triplet superconductor in a Zeeman field. In section III,  we apply this formalism  to describe the phase diagram of a simplest type of fully gapped time-reversal symmetric triplet superconductor, which is described by two independent basis vectors. Then, in section IV we introduce the more general situation, where the triplet superconductor is described by three independent basis vectors.  To elucidate such kind of superconductor, we consider  the so-called Kitaev-Heisenberg model, which can be considered as a paradigmatic model that describes a large variety of different topological phases.   In section V, we  explicitly obtain the various spin-triplet superconducting phases which can be accessed by varying the Zeeman orientation and strength. Finally, in section VI we make some concluding remarks.\\

\section{General Ginzburg-Landau theory}

We consider a spin-triplet superconductor in a external Zeeman-field. The mean-field Bogoliubov--de Gennes Hamiltonian describing the quasiparticles is
\begin{equation}
H_{\textrm{BdG}}(\mathbf{k})=\begin{pmatrix}
h(\mathbf{k})&\Delta(\mathbf{k}) \\
\Delta(\mathbf{k})^\dag & -h(\mathbf{k})^T
\end{pmatrix}, \label{BdG}
\end{equation}
where 
\beq
h(\mathbf{k})=\xi(\mathbf{k}) \sigma_0 + \mathbf{b} \cdot \vec{\sigma}
\eeq 
is the kinetic term,  $\mathbf{b}$ is the Zeeman field, 
\beq
\Delta(\mathbf{k})=i \big(\mathbf{d}(\mathbf{k}) \cdot \vec{\sigma}\big) \sigma_2
\eeq 
is the superconducting order parameter, $\mathbf{d}$ is a three component vector, $\sigma_i$ are the Pauli spin matrices, 
and  we have assumed inversion symmetry such that $\xi(-\mathbf{k})=\xi(\mathbf{k})$. We have neglected the singlet order parameter entirely, as we are working in the regime where the triplet order parameter is energetically favoured. 

The free energy for the Bogoliubov quasiparticles can be written as
\begin{equation}
F_{\textrm{qp}}=\sum_{\mathbf{k}} F_{\textrm{qp}}(\mathbf{k})=-\beta^{-1} \sum_{\mathbf{k}} \sum_{\sigma=\pm} \ln[2\cosh(\beta E_\sigma(\mathbf{k})/2)],
\end{equation}
where $\beta$ is the inverse temperature and the positive quasiparticle energies $E_{\pm}$  can be compactly expressed as
\begin{equation}
E_\pm(\mathbf{k})=\sqrt{\xi(\mathbf{k})^2+|\mathbf{b}|^2+|\mathbf{d}(\mathbf{k})|^2 \pm \sqrt{\lambda(\mathbf{k})}}.
\end{equation}
Here, 
\beq
\bsp
\lambda(\mathbf{k})=&4|\mathbf{b}|^2\xi(\mathbf{k})^2+|\mathbf{q}(\mathbf{k})|^2+4\big(\mathbf{b}\cdot\mathbf{d}(\mathbf{k})\big)\big(\mathbf{b}\cdot\mathbf{d}(\mathbf{k})^*\big)\\
&+4\xi(\mathbf{k}) \mathbf{b}\cdot \mathbf{q}(\mathbf{k}),
\end{split}
\eeq
and 
\beq
\mathbf{q}(\mathbf{k})=i\mathbf{d}(\mathbf{k}) \times \mathbf{d}(\mathbf{k})^*.
\eeq

To describe the effects of a Zeeman field on the superconducting order parameter, we expand the free energy close to the critical temperature $\beta \approx \beta_c$ in the limit  $\beta |\mathbf{b}| \ll 1$ and $\beta |\mathbf{d}| \ll 1$. We obtain
\begin{widetext}
\begin{eqnarray}
F_{\textrm{qp}}(\mathbf{k})&=&F_0(\mathbf{b}, \xi)
-c_1(|\xi|, |\mathbf{b}|^2) |\mathbf{d}|^2-c_2(|\xi|) \bigg\{2  |\mathbf{d}|^4- (\mathbf{d} \cdot \mathbf{d})(\mathbf{d}^* \cdot \mathbf{d}^*)  
+4  (\mathbf{b}\cdot \mathbf{d})(\mathbf{b}\cdot\mathbf{d}^*)
+4  \xi (\mathbf{b} \cdot \mathbf{q}) \bigg\},  \nonumber \\
c_1(|\xi|, |\mathbf{b}|^2)&=&  \frac{\tanh(\beta|\xi|/2)}{2 |\xi|}
+   \frac{(4-\beta^2 |\xi|^2)  \tanh(\beta |\xi|/2)+2 \beta |\xi| \tanh^2(\beta |\xi|/2) + \beta^2 |\xi|^2 \tanh^3(\beta |\xi|/2)-2 \beta |\xi|}{8 |\xi|^3}  |\mathbf{b}|^2, \nonumber
\\
c_2(|\xi|)&=&   \frac{\beta |\xi|-2 \tanh(\beta |\xi|/2)-\beta|\xi| \tanh^2(\beta |\xi|/2)}{16|\xi|^3}\ \ . \label{GLgeneral}
\label{eF}
\end{eqnarray}
\end{widetext}
In addition to the free energy of the Bogoliubov quasiparticles $F_{\textrm{qp}}(\mathbf{k})$, the total free energy also contains a quadratic term in the order parameter $\mathbf{d}(\mathbf{k})$, which appears when one decouples the interaction terms using the  mean field approximation. This term is independent of the Zeeman field, and determines the degenerate solutions of the linearized gap equations. We now express the order parameter as a linear combination of the basis vectors $\mathbf{d}_i$ according to Eq.~(\ref{order_parameter}). By substituting this order parameter into Eqs.~(\ref{GLgeneral}), we obtain an expression for the Ginzburg-Landau free energy per unit cell, given by 
\begin{eqnarray}
\frac{F}{N}&=& \frac{F_0(\mathbf{b})}{N}+[t_I+t(|\mathbf{b}|)] \sum_i |\eta_i|^2 +\kappa_A \sum_i |\eta_i|^4 \nonumber\\&& + 2 \kappa_B \sum_{i < j} |\eta_i|^2 |\eta_j|^2 
%%%%%%%%
-\frac{\kappa_B}{2}\sum_{i<j} (\eta_i^2 \eta_j^{*2}+\eta_j^2 \eta_i^{*2}) \nonumber\\
%%%%%%%
&&
-i D_1 \sum_{i,j,k} \varepsilon_{ijk} b_i \eta_j \eta_k^*  
%%%%%%%%%%%%
+D_2^A \sum_i b_i^2 |\eta_i|^2 \nonumber \\ && -D_2^B \sum_{i<j} b_i b_j (\eta_i \eta_j^*+\eta_j \eta_i^*), \label{Free-energy}
\end{eqnarray}
where the coefficients are given by
\begin{eqnarray}
t(|\mathbf{b}|) &=&-  \sum_{\mathbf{k}} c_1 (|\xi|, |\mathbf{b}|^2)|\mathbf{d}_1|^2/N, \nonumber\\ 
\kappa_A &=&- \sum_{\mathbf{k}} c_2(|\xi|) |\mathbf{d}_1|^4/N, \nonumber\\
\kappa_B &=&-2 \sum_{\mathbf{k}}  c_2(|\xi|) |\mathbf{d}_1|^2  |\mathbf{d}_2|^2/N, \nonumber\\
D_1 &=&4 \sum_{\mathbf{k}} c_2(|\xi|) \xi d_{1,x}  d_{2,y}/ N, \nonumber\\
D_2^A &=&-4 \sum_{\mathbf{k}} c_{2}(|\xi|) |\mathbf{d}_1|^2 /N, \nonumber\\
D_2^B &=&4 \sum_{\mathbf{k}} c_{2}(|\xi|)  d_{1,x}  d_{2,y}/N, \label{coeff}
\end{eqnarray} 
and $c_1 (|\xi|, |\mathbf{b}|^2)$ and $c_2(|\xi|)$ are described in Eq.~(\ref{eF}). Here, $N$ is the number of unit cells, and we have included the term $F_I=N t_I \sum |\eta_i|^2$ arising from the decoupling of the interaction term. Alternatively, the structure of the free energy in Eq.~(\ref{Free-energy}) can be obtained via symmetry arguments.

 \section{Effect of Zeeman field on a time-reversal symmetric helical $p$-wave superconductor with two basis vectors\label{square_lattice}}

We start by considering a simple model for a fully gapped time-reversal symmetric $p$-wave superconductor, which is described by a tight-binding Hamiltonian  on  the square lattice, where the basis vectors corresponding to the largest critical temperature are 
\begin{eqnarray}
\mathbf{d}_1(\mathbf{k})&=&\hat{\mathbf{e}}_x \sin k_y \nonumber\\
\mathbf{d}_2(\mathbf{k})&=&-\hat{\mathbf{e}}_y \sin k_x \ \ , 
\end{eqnarray}
%%%%%%%%%%%
and the single particle dispersion is 
%
%***********************  square lattice dispersion relation ****************
\begin{equation}
\xi=-2t \cos(k_x)-2t \cos(k_y)-\mu 
\label{squarelatticedispersion}
\end{equation}
%**************************************
%
with $t$ denoting the nearest neighbor hopping. This model is equivalent to the models studied in Refs.~\onlinecite{Sato09,Lee12}, and therefore allows us to determine how the self-consistent changes in the superconducting order parameter affect the topological phase transitions predicted there. The Ginzburg-Landau theory for the order parameter $\mathbf{d}(\mathbf{k})=\eta_1 \mathbf{d}_1(\mathbf{k})+\eta_2 \mathbf{d}_2(\mathbf{k})$ is specified by Eq.~(\ref{Free-energy}), where  $D_1=D_2^B=0$  due to symmetries. The other parameters are nonzero and can be evaluated from the expressions (\ref{coeff}). We find that $\kappa_B>0$, which guarantees  that $\eta_1$ and $\eta_2$ can always be chosen to be real. Moreover, we numerically find that $\kappa_A \approx \kappa_B$, and to simplify expressions in what follows, we neglect the small differences between these parameters. Finally, it is useful to measure the Zeeman field and the order parameter relative to a characteristic magnitude of the superconducting order parameter, which we define as $\eta_0^2=-[t_I+t(|\mathbf{b}|)]/3\kappa_A$.  Taking these results into account,  the relevant part of the free-energy is
\begin{eqnarray}
\frac{F}{N}&=& -[t_I+t(|\mathbf{b}|)] \eta_0^2 \bigg\{ \sum_{i=1}^2  \bigg(-1+ \frac{D_2^A}{3 \kappa_A} \frac{b_i^2}{\eta_0^2}\bigg) \frac{\eta_i^2}{\eta_0^2} \nonumber\\&& +\frac{1}{3} \bigg(\frac{\eta_1^4}{\eta_0^4}+\frac{\eta_2^4}{\eta_0^4}+\frac{\eta_1^2\eta_2^2}{\eta_0^4}\bigg) \bigg\}. \label{Free-energy-simplif}
\end{eqnarray}
In the absence of a Zeeman field,  the minimization of the free-energy gives $(\eta_1, \eta_2)=\eta_0 (1, \pm 1)$. The Zeeman field makes the superconductor anisotropic,  so that in the presence of small Zeeman fields the order parameter becomes
\begin{eqnarray}
\eta_1 &=& \eta_0 \sqrt{1-2 \frac{D_2^A}{3 \kappa_A} \frac{b_1^2}{\eta_0^2}+\frac{D_2^A}{3 \kappa_A} \frac{b_2^2}{\eta_0^2}} \ \ , \nonumber \\
\eta_2 &=& \eta_0 \sqrt{1-2 \frac{D_2^A}{3 \kappa_A} \frac{b_2^2}{\eta_0^2}+\frac{D_2^A}{3 \kappa_A} \frac{b_1^2}{\eta_0^2}} \ \ .
\end{eqnarray}
If one of the magnetic field components is sufficiently strong, say $b_1$, then $\eta_1$ becomes zero at the critical magnetic field  
\begin{equation}
 b_1=\frac{\eta_0}{\sqrt{2}}\sqrt{\frac{3\kappa_A}{D_2^A}+ \frac{b_2^2}{\eta_0^2}} \ \ , \label{critb1}
\end{equation}
and for larger Zeeman fields the order parameter is given by
\begin{eqnarray}
\eta_1 &=&0, \nonumber \\
\eta_2 &=& \eta_0 \sqrt{\frac{3}{2}} \sqrt{1- \frac{D_2^A}{3 \kappa_A} \frac{b_2^2}{\eta_0^2}} \ \ .
\end{eqnarray}
Numerically, we find that $3 \kappa_A/D_2^A \sim 1$, which means that the critical field at which one of the components of the order parameter vanishes is on the order of $\eta_0$. We  note that although $b_3$ does not explicitly appear in the equations above, it still influences the order parameter, because the parameter $\eta_0$ depends on $|\mathbf{b}|$.

We now  consider the case studied in Ref.~\onlinecite{Lee12}, where the Zeeman field is in the (x,y)-plane. In this case, it was found \cite{Lee12} that the Hamiltonian satisfies a nontrivial chiral symmetry
\begin{equation}
S_1^\dag H(\mathbf{k})S_1=- H(\mathbf{k}) \ \ ,
\end{equation}
where 
\begin{equation}
S_1=\sigma_1 \otimes \sigma_1\ \ . \label{S1}
\end{equation}
The chiral symmetry allows to rewrite the Hamiltonian  in a block-off-diagonal form
\begin{equation}
U_1^\dag H(\mathbf{k}) U_1=\begin{pmatrix}
0 & A_1(\mathbf{k}) \\
A_1(\mathbf{k})^\dag & 0
\end{pmatrix}, \label{off-diag}
\end{equation}
where in this case $U_1=(\sigma_1 \otimes \sigma_1+\sigma_3 \otimes \sigma_0)/\sqrt{2}$ and
\begin{equation}
A_1(\mathbf{k})=h(\mathbf{k})\sigma_1-\Delta(\mathbf{k}) \ \ . \label{A1}
\end{equation}
Furthermore, it is easy to see from Eq.~(\ref{off-diag}) that the quasiparticle spectrum has nodes if the  equations
\begin{subequations}
\begin{eqnarray}
|\mathbf{b}|^2&=&\xi(\mathbf{k})^2+|\mathbf{d}(\mathbf{k})|^2  \label{firstcondition.eq}\\
b_1 d_2(\mathbf{k})&=&b_2 d_1(\mathbf{k}) \label{secondcondition.eq}
\end{eqnarray}
\end{subequations}
are simultaneously satisfied. Since the relative magnitude  of the two  $\mathbf{d}$-vector components changes as a function of momentum direction, $d_1(\mathbf{k})=\eta_1 \sin(k_y)$ and  $d_2(\mathbf{k})=-\eta_2 \sin(k_x)$, the second equation is always satisfied for specific directions in  momentum space. In particular, there exists a momentum $\mathbf{k}_0$ at the Fermi surface $\xi(\mathbf{k}_0)=0$ such that $b_1 d_2(\mathbf{k}_0)=b_2 d_1(\mathbf{k}_0)$. This momentum $\mathbf{k}_0$ defines the smallest 
Zeeman field $b_{c, \rm{ch}}$  for which the first condition can be satisfied. For larger Zeeman fields  
\begin{equation}
|\mathbf{b}|> b_{c, \rm{ch}} \equiv  |\mathbf{d}(\mathbf{k}_0)| \ \ , 
\end{equation}
both conditions can be satisfied simultaneously due to the fact that there is actually a one-parameter family of vectors $\mathbf{k}$ which satisfies Eq.~(\ref{secondcondition.eq})
but has a nonzero $\xi(\mathbf{k})$. 
Therefore, for a sufficiently strong Zeeman field in the $(x,y)$-plane, $|\mathbf{b}|>b_{c, \rm{ch}} \sim \eta_0$, we  expect a nodal phase with a chiral symmetry. This phase exists even if the order parameter is not calculated self-consistently as found in Ref.~\onlinecite{Lee12}, and the anisotropy of the order parameter affects $b_{c, \rm{ch}}$ only quantitatively. 

The importance of a self-consistent computation of the order parameter becomes apparent when one considers a Zeeman field in the $(x,z)$-plane. Then, the chiral symmetry defined above no longer exists. However, a new type of chiral symmetry 
\begin{equation}
S_2=\sigma_1 \otimes \sigma_0 \label{S2}
\end{equation}
 emerges when $b_1$ is sufficiently large so that $\eta_1=0$. The Hamiltonian 
can be written in a block-off-diagonal form, similarly as in Eq.~(\ref{off-diag}), where now $U_2=(\sigma_0 \otimes \sigma_0+i \sigma_2 \otimes \sigma_0)/\sqrt{2}$ and
\begin{equation}
A_2(\mathbf{k})=h(\mathbf{k})+\Delta(\mathbf{k}). \label{A2}
\end{equation}
Now $\Delta(0, k_y)=0$ for all $k_y$, and therefore nodes appear in the quasiparticle spectrum at $\xi(0,k_y)=\pm |\mathbf{b}|$. Although the mechanism giving rise to a nodal phase with chiral symmetry is now different, the critical field is on same order of magnitude  $b_{c, \rm{ch}}=\frac{\eta_0}{\sqrt{2}}\sqrt{\frac{3\kappa_A}{D_2^A}} \sim \eta_0$.

Now that we have discussed the behavior of the order parameter in a Zeeman field, we are ready to discuss the different topological phases as a function of the Zeeman field magnitude and the orientation. 

\subsection{Time-reversal broken topological superconductivity}

\begin{figure}[tbp]
\centering\includegraphics[width=8.5cm]{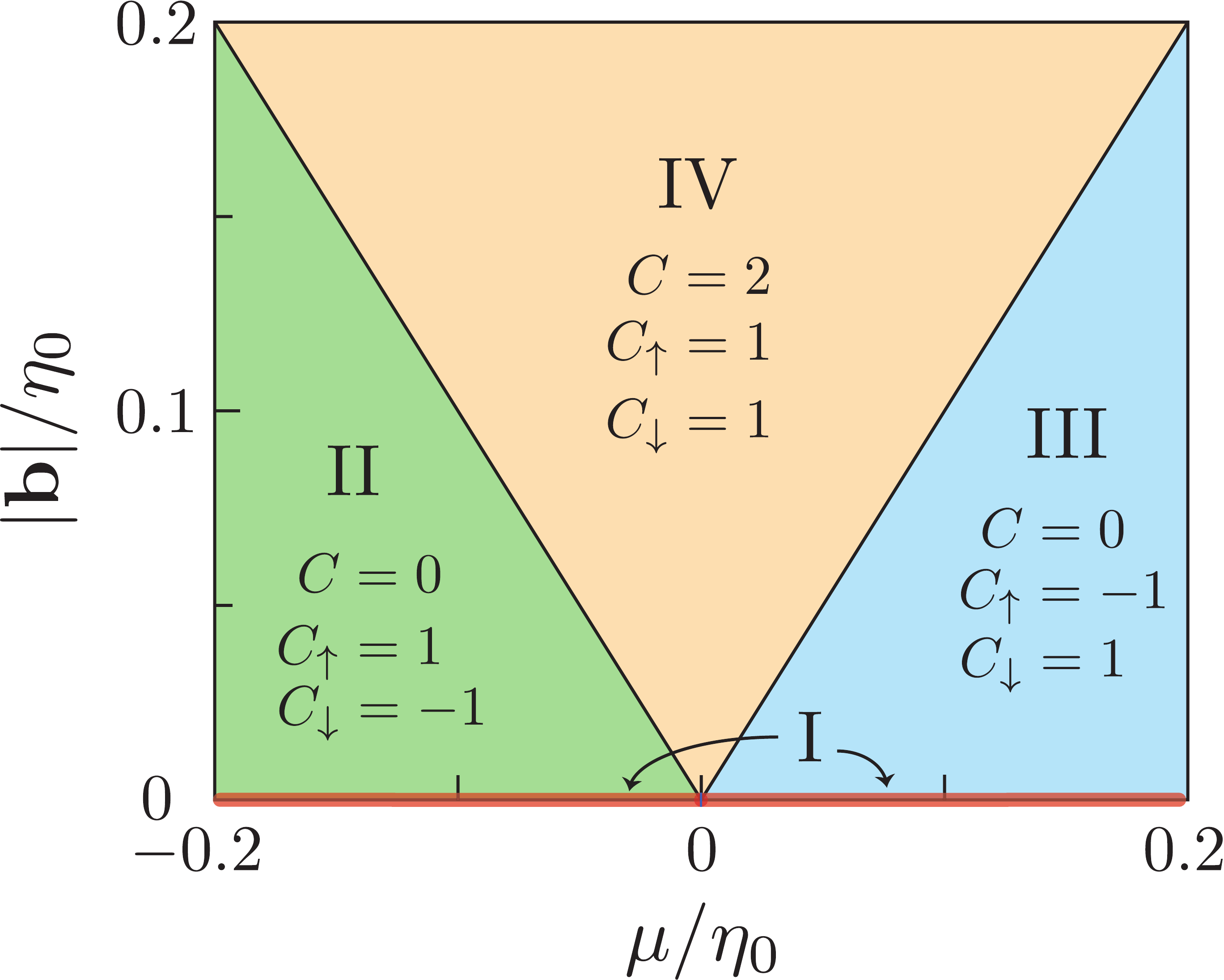}
\caption{Phase diagram of the helical $p$-wave superconductor with two basis vectors near $\mu= 0$ as a function of Zeeman field $|\mathbf{b}|$. The four phases are I: time reversal invariant topologically nontrivial $\mathbb{Z}_2$ superconducting phase with counterpropagating edge modes, II $\&$ III: time reversal broken  $p$-wave with $U_s$-symmetry protected edge modes, IV: time reversal broken topologically non-trivial p-wave superconducting phase. In phases II and III,  a Zeeman field rotated away from the $(0, 0, 1)$ direction will break the symmetry protection and hybridise the edge modes.}
\label{phasefigsquare}
\end{figure}

%Description of the phases

We start by reviewing the various fully gapped topological phases of the model, which were already discussed in Ref.~\onlinecite{Sato09}. Below, we show that the existence of these phases does not depend on the isotropy of the order parameter, and thus there are no qualitative changes due to  the self-consistent computation of the order parameter in the presence of a Zeeman field. 

In order to understand where  phase transitions between topologically distinct fully gapped phases can take place, we first note that in addition  to our earlier considerations the energy gap can also close at the high symmetry points in the Brillouin zone. 
The conditions for these gap closings are  
\begin{eqnarray}
|\mathbf{d}(\mathbf{K}_i)|=0,  \ \ \ \xi(\mathbf{K}_i)=\pm |\mathbf{b}|, \label{nodesK}
\end{eqnarray}
and the possible high-symmetry points $\mathbf{K}_i$ are $\Gamma=(0,0)$, $M_1=(\pi,0)$, $M_2=(0, \pi)$ and $M_3=(\pi, \pi)$. The nature of these gap closing is very different from the gap closings discussed before, because here the energy gap closes only at a specific value of the Zeeman field, and after it reopens the topological invariant describing the number of the protected edge modes has  changed.

To describe these phase-transitions, we can for simplicity assume that the Zeeman field  points in the $z$-direction, so that  the fermions with different spins are completely decoupled,  corresponding to an additional symmetry $[H, U_s]=0$, where $U_s=\sigma_0 \otimes \sigma_3$. Then, the Hamiltonian can be written as  
\begin{equation}
U^\dag H(\mathbf{k}) U=\begin{pmatrix}
H_\uparrow(\mathbf{k}) & 0 \\
0 & H_\downarrow(\mathbf{k})
\end{pmatrix},
\end{equation}
where $H_\uparrow(\mathbf{k})= [\xi(\mathbf{k})+b_3] \sigma_3-\eta_1 \sin k_y  \sigma_1+\eta_2 \sin k_x  \sigma_2$ and $H_\downarrow(\mathbf{k})= [\xi(\mathbf{k})-b_3] \sigma_3+\eta_1 \sin k_y  \sigma_1+\eta_2 \sin k_x  \sigma_2$.
The block diagonal form of the Hamiltonian obtained this way is useful for  understanding  all the different topological phases, although it is not necessary for the existence of the  strong topological invariants discussed below. The different topological phases can now be found by computing the Chern number for each spin-block
\begin{equation}
C_{\uparrow, (\downarrow)} = \sum_{n \in\mathrm{filled}}\frac{1}{\pi}\int_{BZ} d^2 k \, \mathrm{Im}\bigl(\langle \partial_{k_x} \psi_{n, \uparrow (\downarrow) \mathbf{k}}|\partial_{k_y}\psi_{n,\uparrow (\downarrow) \mathbf{k}}\rangle\bigr). \label{Chern_number}
\end{equation}
In the time-reversal broken case, the strong $\mathbb{Z}$ topological invariant is given by the total Chern number $C \equiv C_{\uparrow}+C_{\downarrow}$.
For time-reversal invariant superconductors, the total Chern number $C \equiv C_{\uparrow}+C_{\downarrow}=0$, and the $\mathbb{Z}_2$ invariant $\nu$ is determined by the parity of $C_{\uparrow}$ \cite{comment2}. In addition to these strong topological indices, one can define two $U_s$-symmetry-protected  $\mathbb{Z}$ topological invariants $C_\uparrow$ and $C_\downarrow$. These invariants are "weaker" than the strong topological indices discussed above, and guarantee the existence of  edge modes only in the presence of the symmetry $U_s$.

The most interesting topological phase transitions occur at  carrier densities for which the conditions (\ref{nodesK}) are satisfied at the $M_1$ and $M_2$ points. While $|\mathbf{d}(\mathbf{K}_i)|=0$ is automatically satisfied at these points, the second condition requires that $\mu=\pm |\mathbf{b}|$. We find that $C_{\uparrow}=1$ for $\mu<|\mathbf{b}|$ and $C_{\uparrow}=-1$ for $\mu>|\mathbf{b}|$, whereas $C_{\downarrow}=-1$  for $\mu<-|\mathbf{b}|$ and $C_{\downarrow}=1$ for $\mu>-|\mathbf{b}|$. This way, we arrive at the phase-diagram shown in Fig.~\ref{phasefigsquare}. Most importantly, by choosing the chemical potential in the vicinity of the topological phase transition $\mu=0$, the Zeeman splitting effectively pushes the chemical potential of one spin species into the electron doped regime, and the other into the hole doped regime, so that $C =2$ (region IV). Once  $C \ne 0$, the block diagonal form of the Hamiltonian is no longer essential, and the edge modes are robust against perturbations  which couple the spin blocks. Thus, this topologically nontrivial phase which supports two topologically protected chiral Majorana edge modes, exists for a wide range of magnitudes and directions of the Zeeman field. Additionally, the phase diagram in Fig.~\ref{phasefigsquare} contains a time-reversal invariant topologically nontrivial phase at $|\mathbf{b}|=0$ (region I), which supports counterpropagating edge modes that are protected by the time-reversal symmetry. For completeness, we have also included there the $U_s$-symmetry protected topological phases (regions II and III), which are distinguished by topological numbers $C_\uparrow$ and $C_{\downarrow}$ if the Zeeman field points along $z$-axis.

Topological phase transitions can also be found near the $\Gamma$ ($\mu = -4t \pm |\mathbf{b}|$) and $M_3$ ($\mu=4t \pm |\mathbf{b}|$) points \cite{Sato09}. However, these phase transitions take place close to the band edges, and therefore one expects that the critical temperature for the superconductivity and the magnitude of the superconducting order parameter $\eta_0$ are much smaller. Thus, these transitions are less interesting from an experimental point of view.

\subsection{Majorana flat bands}

\begin{figure}
\includegraphics[scale=0.21]{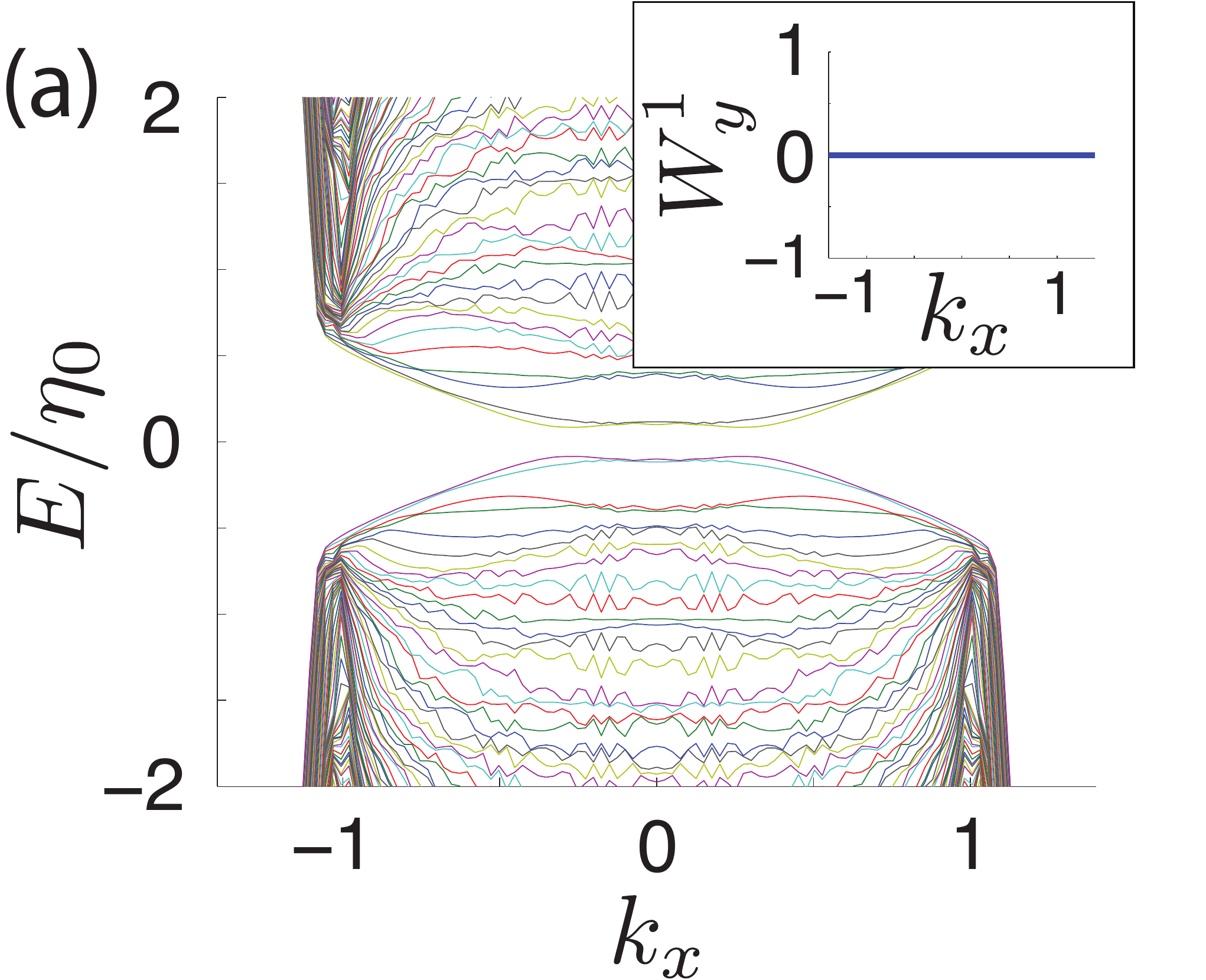}\includegraphics[scale=0.21]{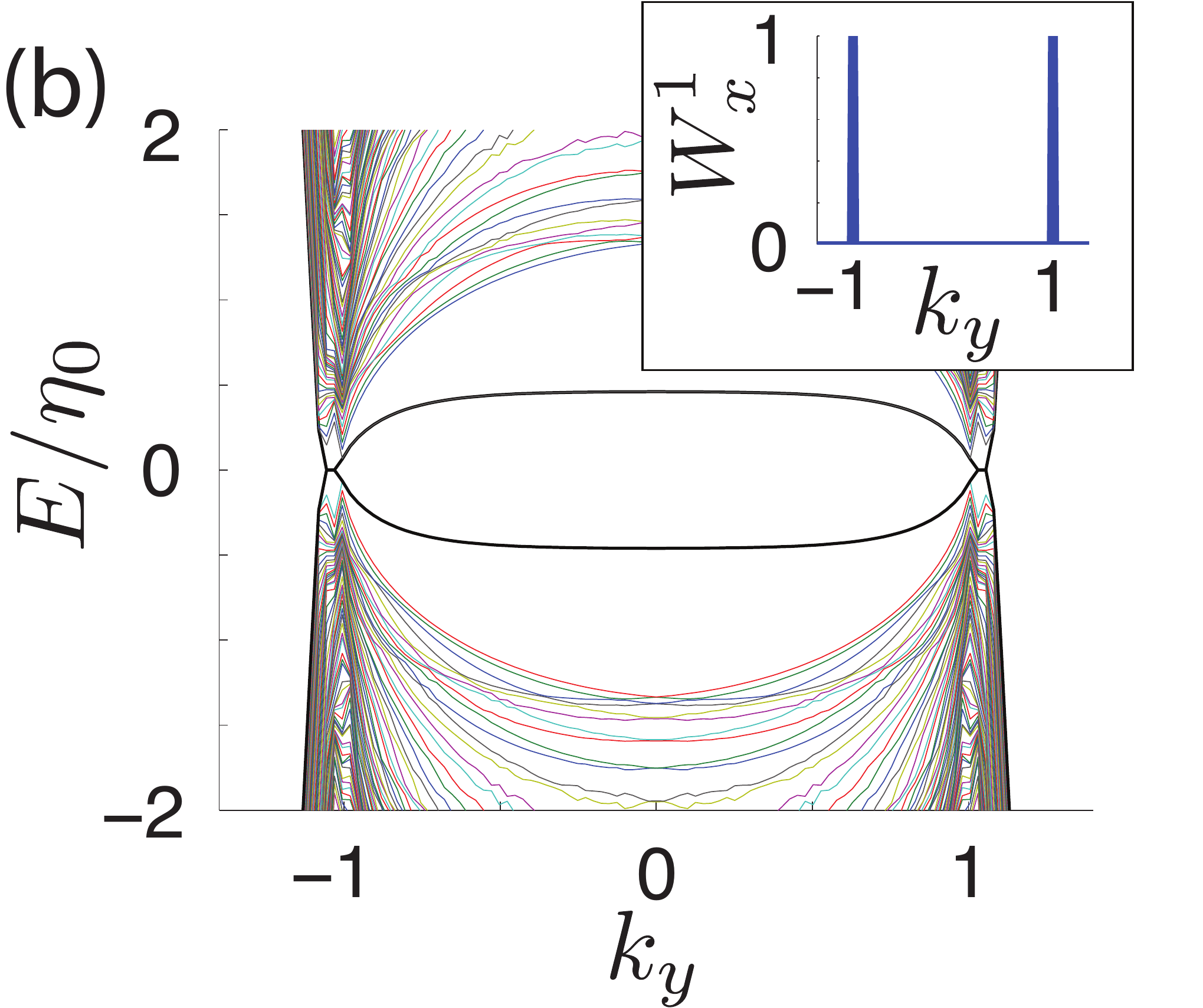}
\caption{Helical $p$-wave superconductor with two basis vectors: Majorana flat bands and the corresponding topological invariants for edges along (a) $x$- and (b) $y$-directions and a Zeeman field $b_1=0.8 \eta_0$, $b_2=0.6 \eta_0$, $b_3=0$. In Fig.~(b) Majorana flat bands appear in a small intervals around $k_y=\pm k_F \approx \pm 1$. The order parameter was calculated self-consistently yielding  $\eta_1=0.28 \eta_0$, $\eta_2=0.96 \eta_0$. However, qualitatively similar results can be obtained also if the self-consistent changes are not taken into account.}
\label{flatbandsxyplane}
\end{figure}

 It turns out that the nodal superconducting phases with a chiral symmetry ($|\mathbf{b}|>b_{c, \rm{ch}}$) can support Majorana flat bands.
To understand these flat bands, we consider a translationally invariant system in one direction, say the $x$-direction. This means that $k_x$ is a good quantum number, and that for each value of $k_x$ we have a one-dimensional Hamiltonian $H_{k_x}(k_y)$ which depends only on $k_y$. These one-dimensional Hamiltonians $H_{k_x}(k_y)$ are gapped whenever there are no nodes for any value of $k_y$, and thus one-dimensional topological invariants are well-defined. Suppose there are two nodal points across the two-dimensional Brillouin zone, at $k_x^1$ and $k_x^2$. They correspond to two nodal  Hamiltonians, $H_{k_x^1}(k_y)$, and $H_{k_x^2}(k_y)$.  If for some value of $k'_x$, where $k_x^1<k'_x<k_x^2$, for instance, the one-dimensional Hamiltonian $H_{k'_x}(k_y)$ yields a non-trivial topological number, the same topological number must occur for all $k_x$ across the entire interval $(k_x^1,k_x^2)$, because these topological numbers can change only when the energy gap closes at nodal points. Because nontrivial topological number gives rise to a zero-energy edge state, then each of these one-dimensional Hamiltonians supports a zero energy edge state. The collection of these edge states between the two nodal Hamiltonians then naturally leads to topological Majorana flat bands.

To be more specific, by using the block-off-diagonal form of the Hamiltonian, Eq.~(\ref{off-diag}),  we can define $\mathbb{Z}$-topological invariants (winding numbers) \cite{Manmana12, Tewari,Lee12}
\begin{eqnarray}
W_x^{i}(k_y)=\frac{-i}{2 \pi} \int dk_x \frac{1}{z_{i}}\frac{dz_{i}}{dk_x}, \nonumber \\
W_y^{i}(k_x)=\frac{-i}{2 \pi} \int dk_y \frac{1}{z_{i}}\frac{dz_{i}}{dk_y}, \label{windingnumber}
\end{eqnarray}
where the integration is over the one-dimensional Brillouin zone and
\begin{equation}
z_{i}(\mathbf{k})=\frac{\det[A_{i}(\mathbf{k})]}{|\det[A_{i}(\mathbf{k})]|}.
\end{equation}
The topological invariant $W_x^i(k_y)$ describes the number of Majorana flat bands in a system with periodic boundary conditions in $y$-direction and open boundary conditions in $x$-direction, whereas $W_y^i(k_x)$ describes the number of Majorana flat bands when the boundary conditions for $x$- and $y$-directions are exchanged. Here, the  index $i$ labels the possible chiral symmetries $S_i$ [Eqs.~(\ref{S1}) and (\ref{S2})] and the corresponding matrices $A_i$ [Eqs.~(\ref{A1}) and (\ref{A2})]. 

\begin{figure}
\includegraphics[scale=0.21]{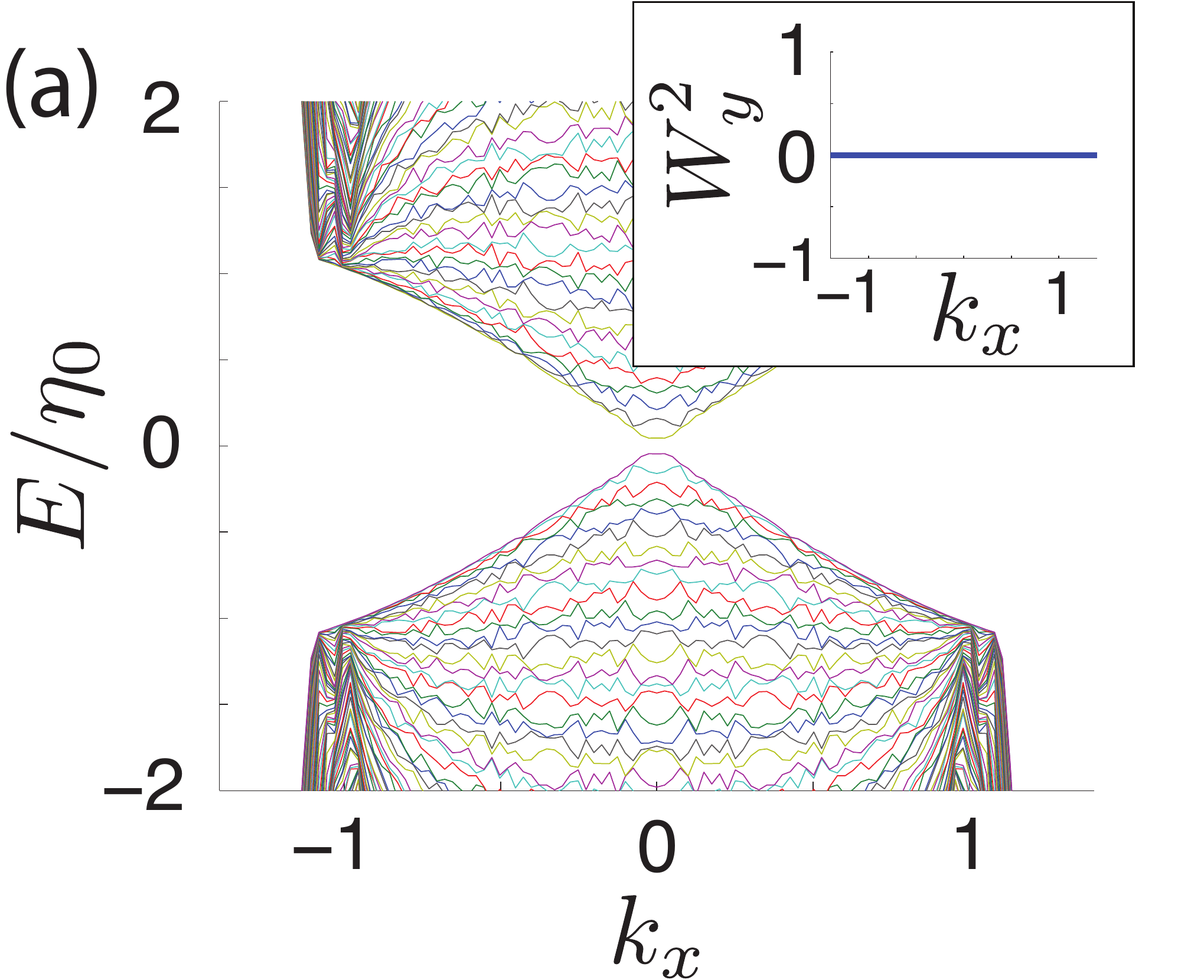}\includegraphics[scale=0.21]{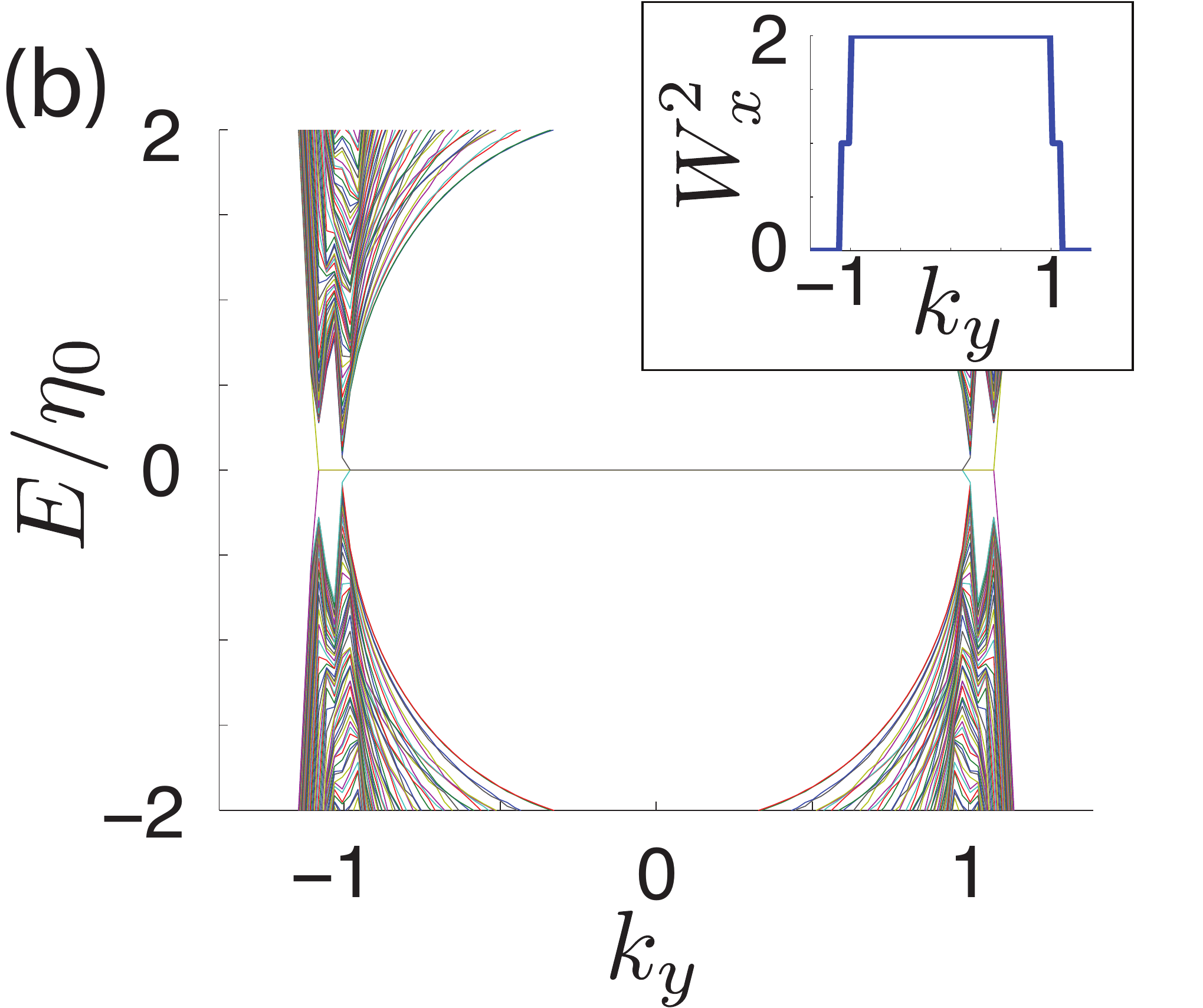}
\includegraphics[scale=0.21]{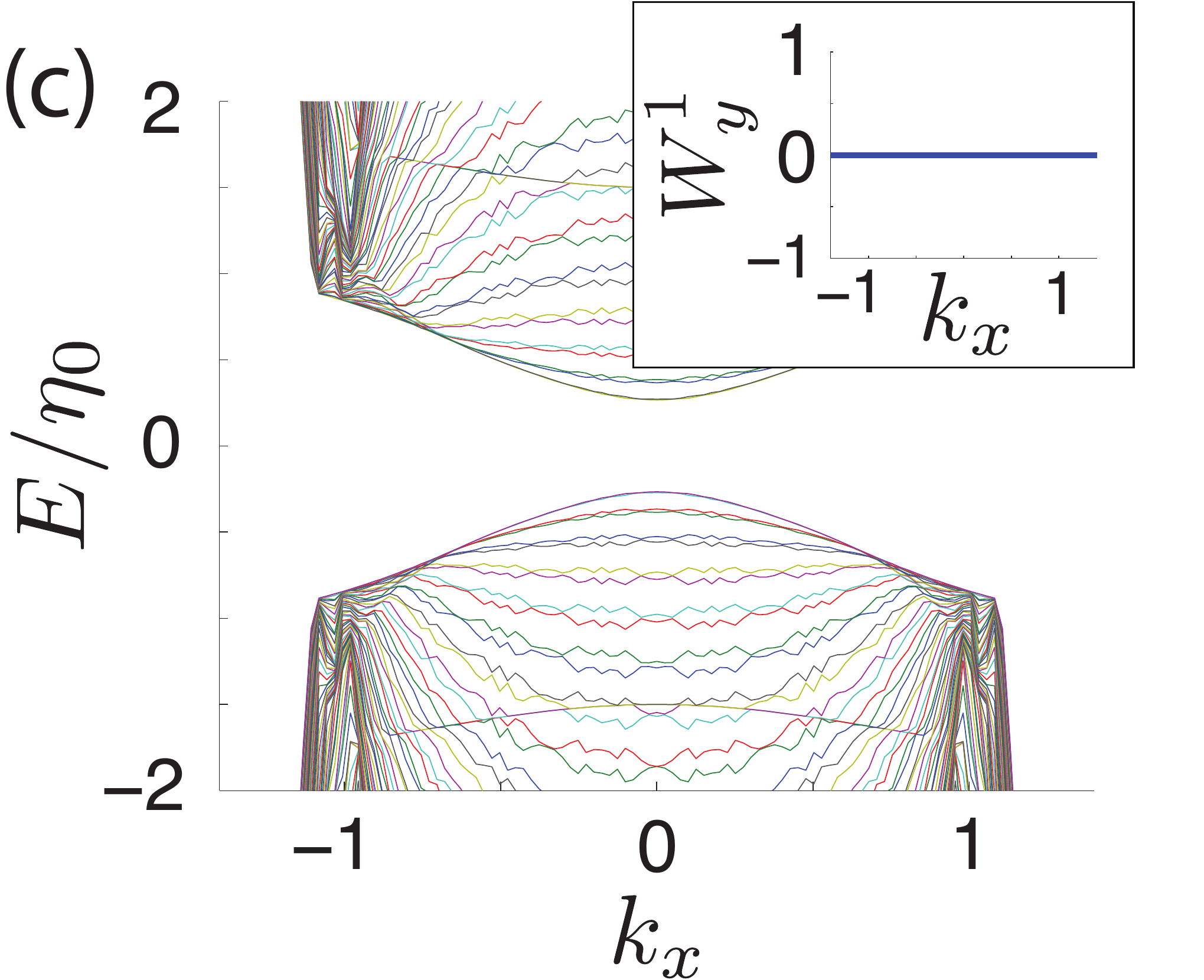}\includegraphics[scale=0.21]{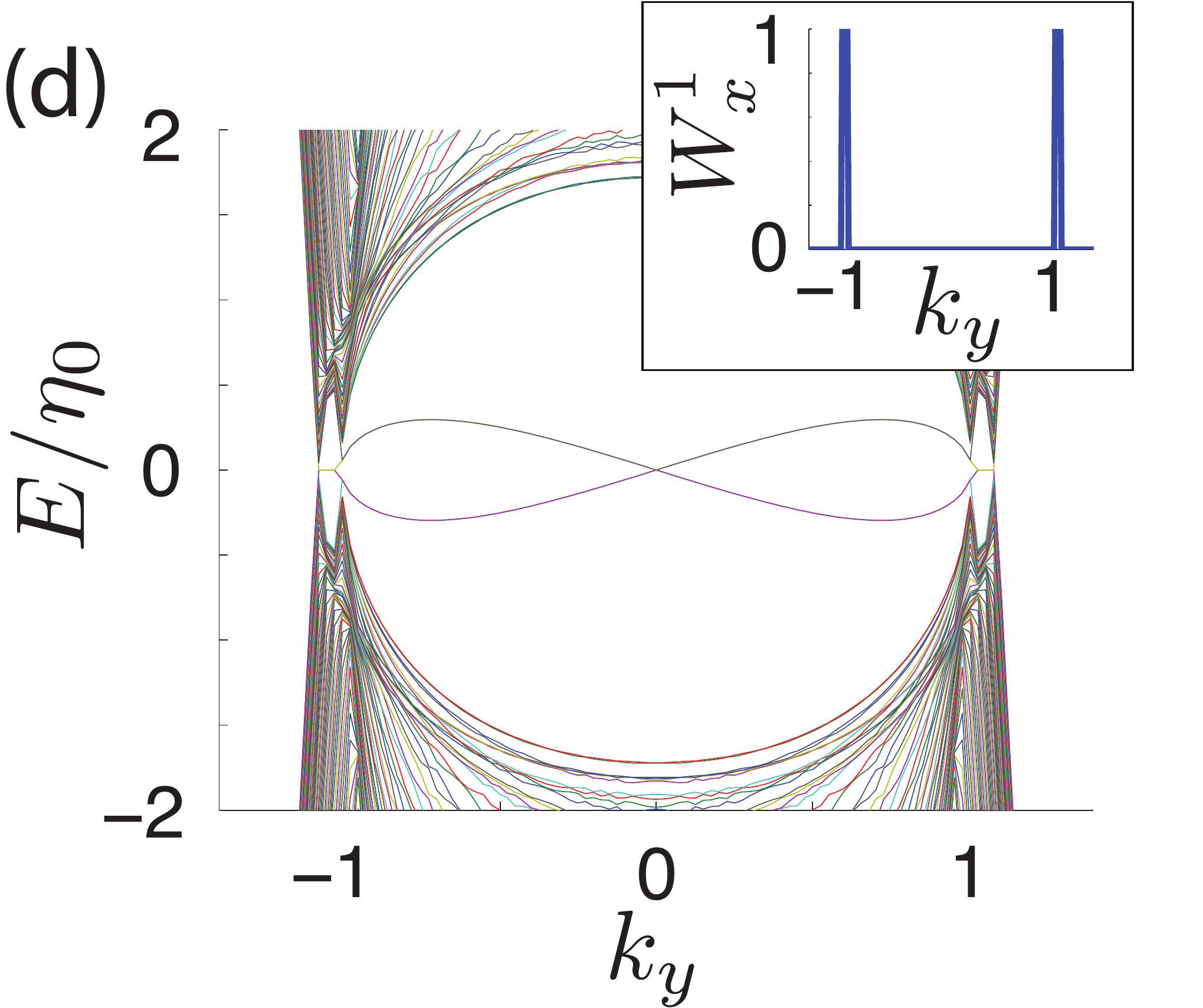}
\caption{Helical $p$-wave superconductor with two basis vectors: Majorana flat bands for a Zeeman field in $x$-direction with $b_1=1.5 \eta_0$, $b_2=b_3=0$.
(a), (b) Majorana flat bands and the corresponding topological invariants for self-consistently calculated order parameter,  $\eta_1=0$ and $\eta_2=\eta_0 \sqrt{3/2}$, for edges along $x$- and $y$-directions, respectively.
(c), (d) Majorana flat bands and the corresponding topological invariants for non-self-consistent order parameter, $\eta_1=\eta_0$ and $\eta_2=\eta_0$.}
\label{flatbandsxaxis}
\end{figure}

\begin{figure}
\includegraphics[scale=0.21]{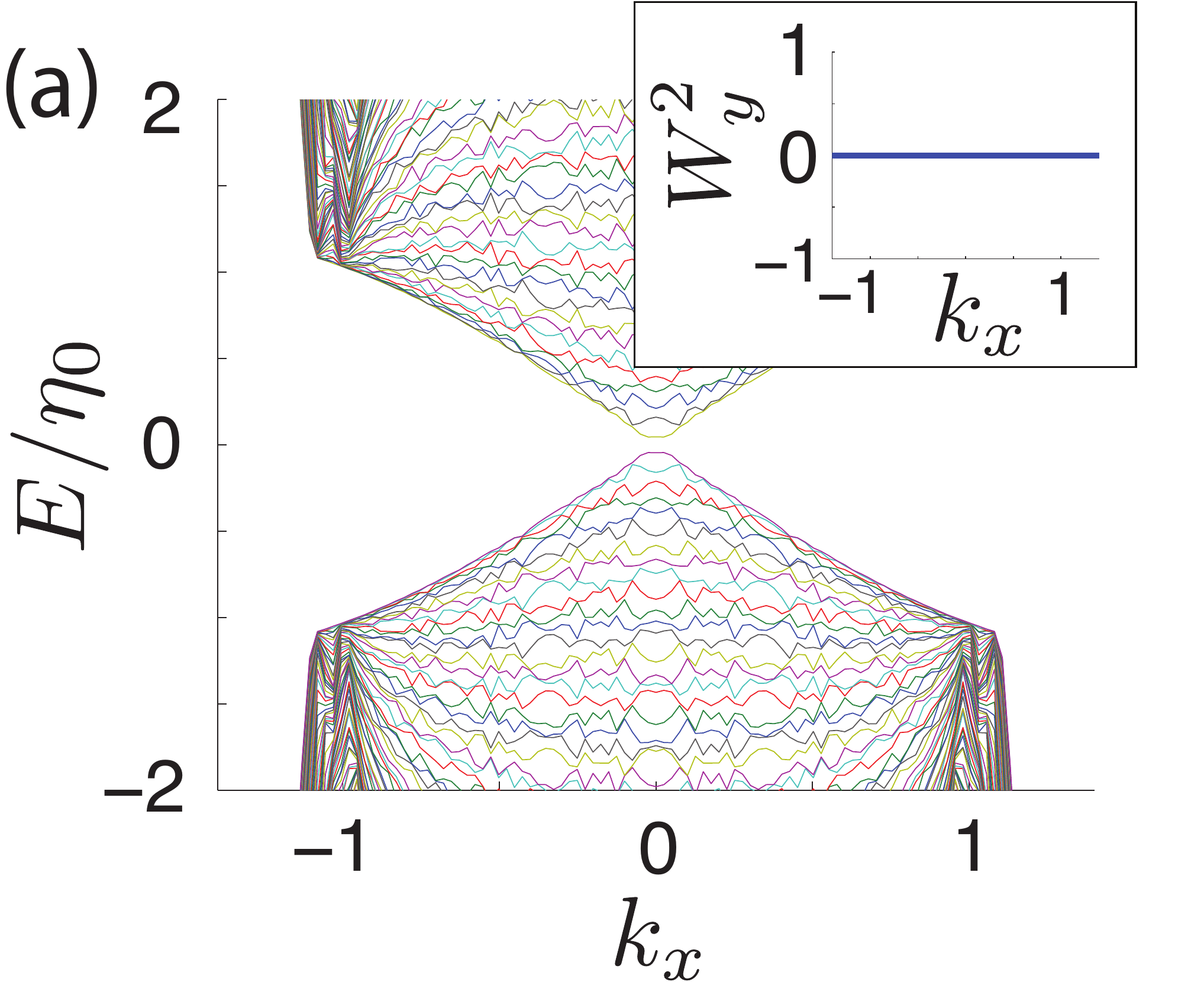}\includegraphics[scale=0.21]{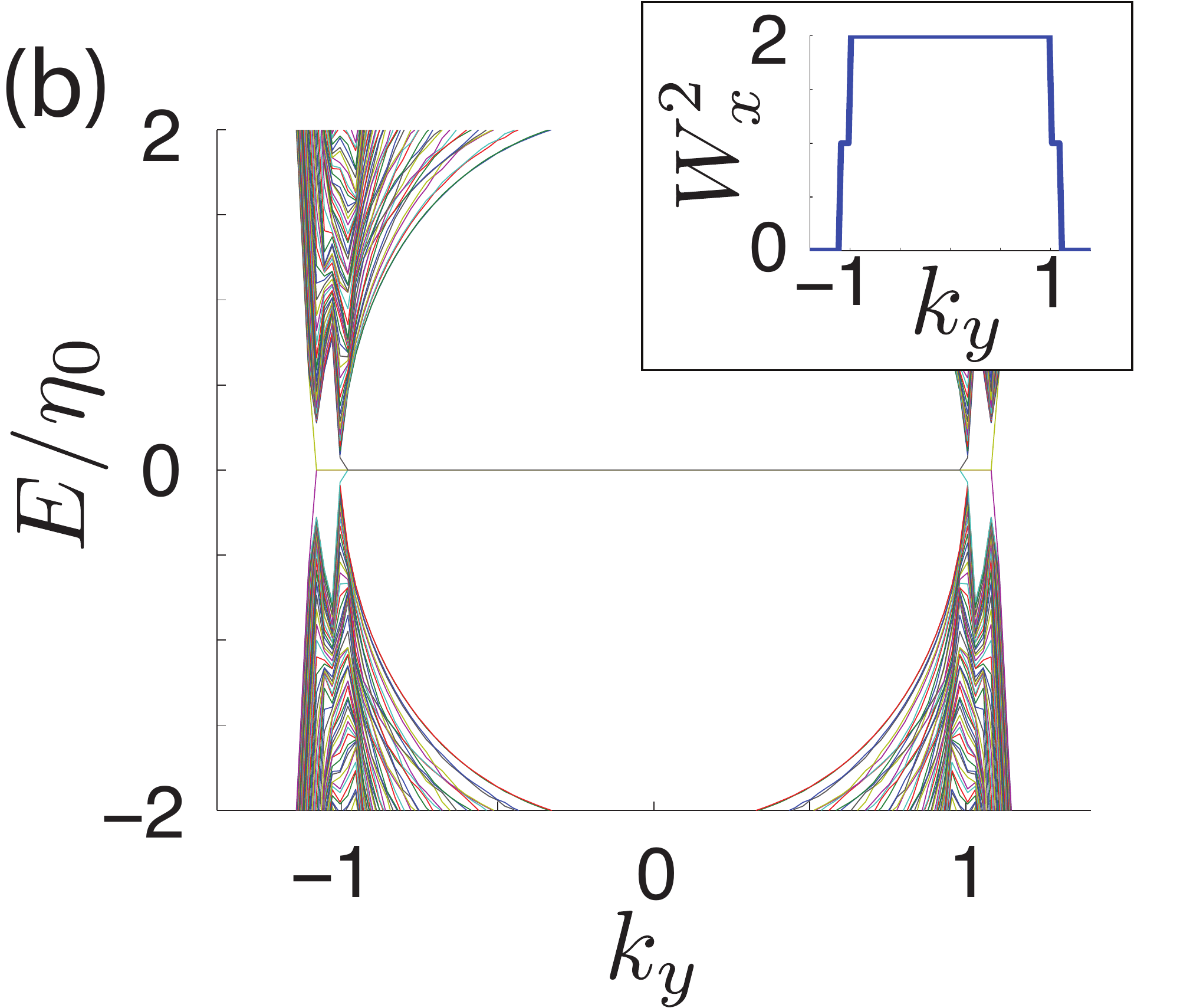}
\includegraphics[scale=0.21]{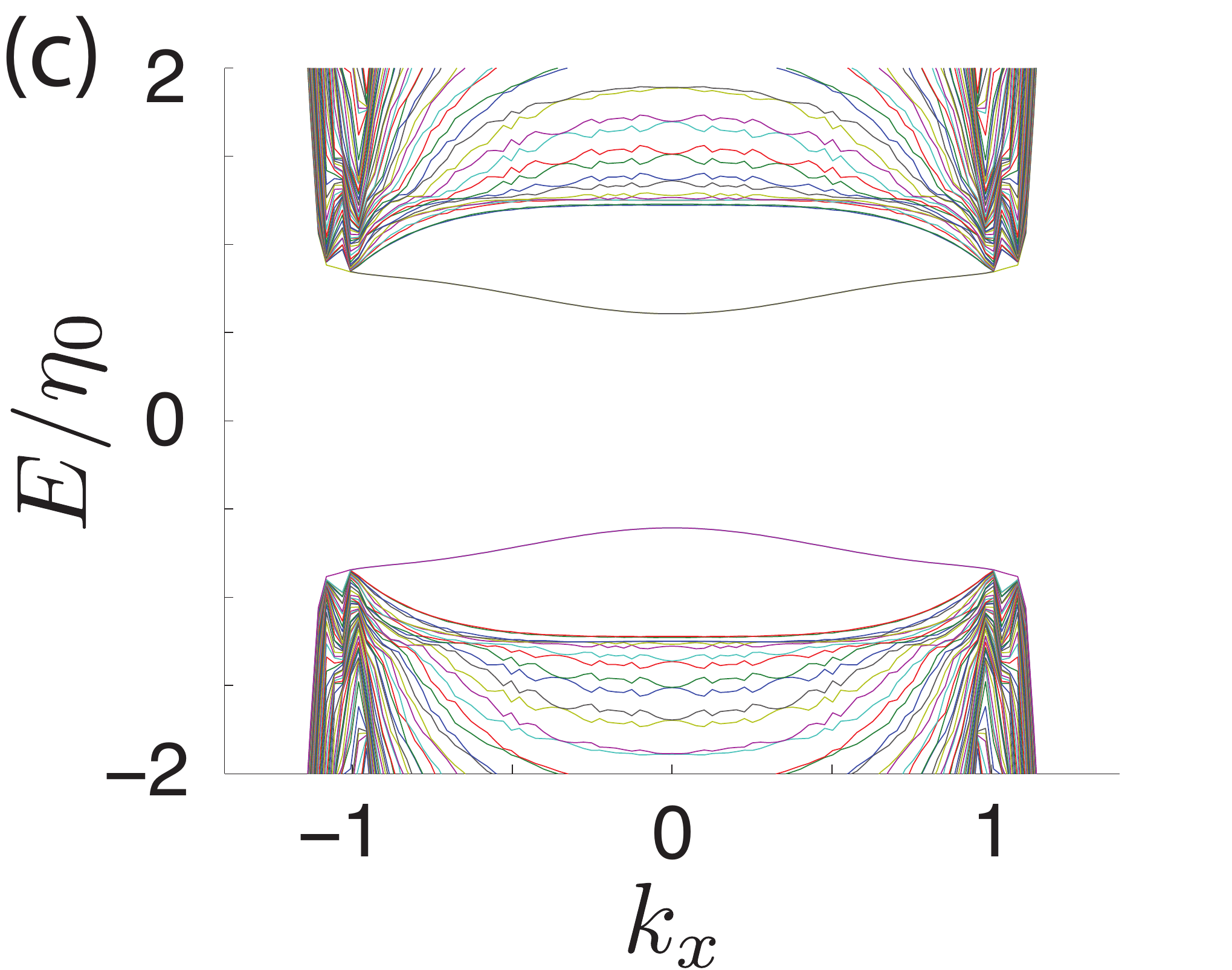}\includegraphics[scale=0.21]{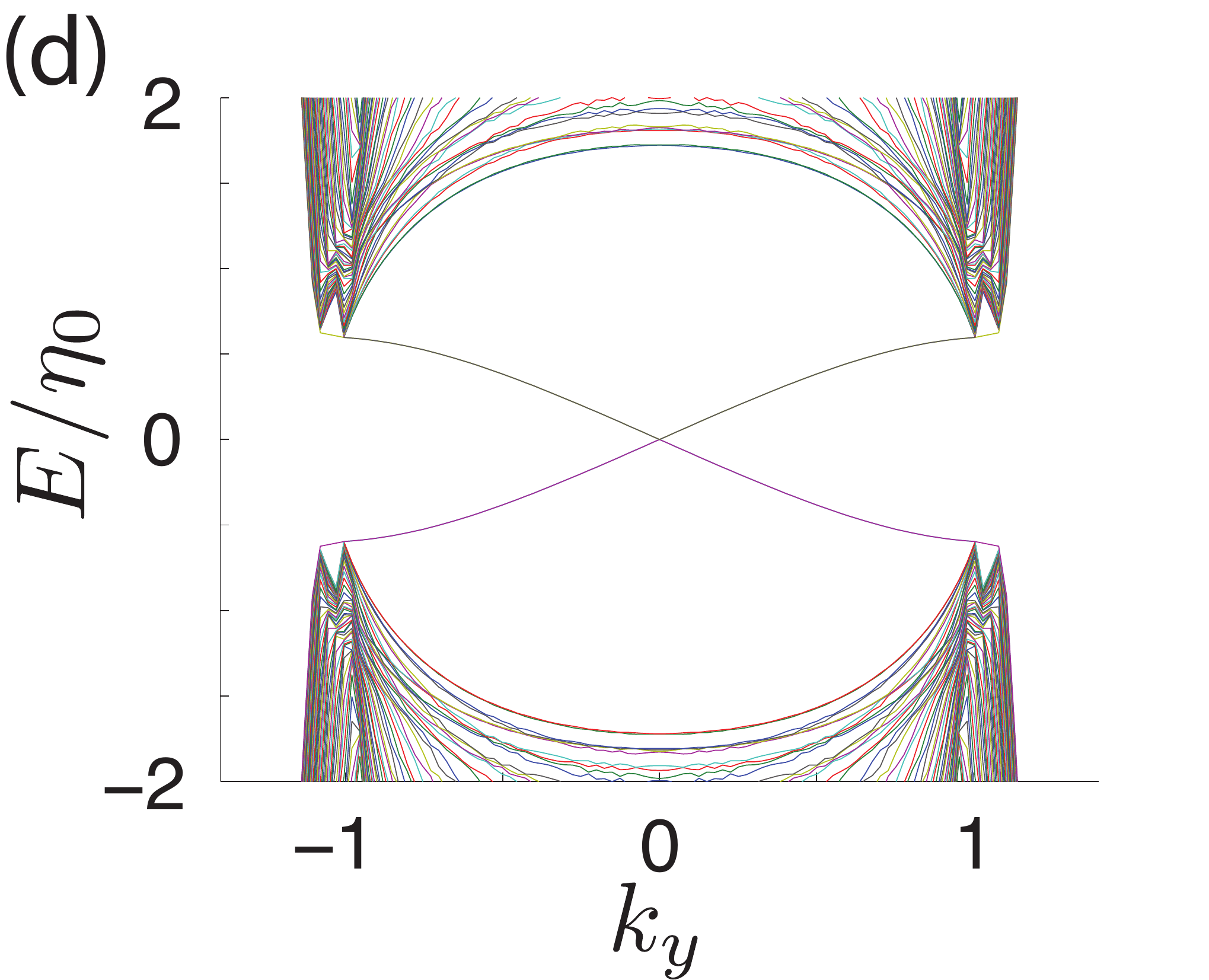}
\caption{Same as Fig.~\ref{flatbandsxaxis} but for $b_1=b_3=1.5 \eta_0/\sqrt{2}$, $b_2=0$. }
\label{flatbandszxplane}
\end{figure}

The topological invariant and flat bands are illustrated in Figs.~\ref{flatbandsxyplane}, \ref{flatbandsxaxis} and \ref{flatbandszxplane} for several different Zeeman field directions. In order to compute the self-consistent changes in the order parameter, we assume that  $3 \kappa_A/D_2^A = 1$, which is close to the numerical value obtained from the expressions Eq.~(\ref{coeff}). Additionally,  we choose $\eta_0=0.05 t$ and $\mu=-3t$ for all numerical results. 
Fig.~\ref{flatbandsxyplane} illustrates the situation for a generic Zeeman field direction in the $(x,y)$-plane such that both $b_1, b_2 \ne 0$.  Then, the relevant chiral symmetry is $S_1$, and a possible anisotropy of the order parameter is not important, so that qualitatively similar results are obtained both for self-consistently and non-self-consistently calculated order parameters. Thus, the results are in good agreement with the predictions of Ref.~\onlinecite{Lee12}, and Majorana flat bands appear between the nodes corresponding to the different spin bands around $k=k_F$. Because the magnitude of the Zeeman field cannot be much larger than $\eta_0(|\mathbf{b}|=0)$ without  destroying the superconductivity,  the interval of the flat bands in the momentum space has a maximum size $\Delta k \sim (\eta_0(|\mathbf{b}|=0)/E_F) k_F$  [See Fig.~\ref{flatbandsxyplane} (b)]. Here, $k_F$ and $E_F$ are the Fermi wave vector and Fermi energy (with respect to the band bottom), correspondingly. Thus, in weak coupling theories of superconductivity ($\eta_0(|\mathbf{b}|=0) \ll E_F$),  $\Delta k$ is always quite small as compared  to $k_F$.  

\begin{table*}
  \caption{Summary of the known topologically nontrivial phases of the Kitaev-Heisenberg model.}
  \label{tab:top_phases}
  %\begin{ruledtabular}
  \begin{tabular}{| c | c | c | }
\hline
  \emph{Spin liquid phases (SL)} & \emph{Parameter regime} & \emph{Ref.}\\
  \hline
%  \hline
    Gapless SL with flat bands &  Undoped, $J_K$ dominates,  $\mathbf{b}=0$ & \onlinecite{Kitaev06}\\
    Fully gapped topological SL&  Undoped, $J_K$ dominates, $b_x b_y b_z \ne 0$ & \onlinecite{Kitaev06} \\
    \hline\hline
   \emph{Superconducting phases (SC)}  & \emph{Parameter regime} & \emph{Ref.} \\
    \hline
 %   \hline
    Fully gapped time-reversal broken $d$-wave SC & $J_H>0$ dominates, $\mathbf{b}=0$ & \onlinecite{BlSc07}\\
    Fully gapped time-reversal broken topological $p$-wave SC & Small doping, $J_K>0$ dominates, $\mathbf{b}=0$ &\onlinecite{Vishwanath} \\
    Fully gapped time-reversal symmetric topological $p$-wave SC & Large doping, $J_K>0$ dominates, $\mathbf{b}=0$ & \onlinecite{Hyart12}\\
    Fully gapped $p$-wave SC with symmetry-protected edge modes&  Intermediate doping, $J_K>0$ dominates, $\mathbf{b}=0$ & \onlinecite{Hyart12}\\
    Fully gapped time-reversal broken topological SC & $|\mu| \approx t$, $J_K>0$ dominates, $\mathbf{b} \ne 0$ & Here \\
    Nodal p-wave SC with flat bands & $J_K>0$ dominates, $\mathbf{b}\parallel (x,y)-, \ (y,z)-, \ (z,x)-$planes & Here\\
    \hline
  \end{tabular}
%   \end{ruledtabular}
\end{table*}

Self-consistent changes in the order parameter  have dramatic effects once the second type of chiral symmetry $S_2$ emerges, as demonstrated in Figs.~\ref{flatbandsxaxis} and \ref{flatbandszxplane}. Fig.~\ref{flatbandsxaxis} shows the Majorana flat bands for a sufficiently strong Zeeman field along $(1,0,0)$ direction such that  $\eta_1=0$ and the chiral symmetry $S_2$ is satisfied. As can be seen in Fig.~\ref{flatbandsxaxis}(b) the flat bands now exist in the whole interval of $k_y$ from $-k_F$ to $k_F$. We stress that this result cannot be explained with the model used in Ref.~\onlinecite{Lee12}. In contrast, if the self-consistent changes in the order parameter are not taken into account, $S_2$ is not valid, and the chiral symmetry $S_1$ gives rise to a flat band only in the small interval $\Delta k  \sim (\eta_0/E_F) k_F$ around $k=k_F$ [Fig.~\ref{flatbandsxaxis}(d)]. We also point out that the flat bands remain unaffected if the Zeeman field is rotated away from the $(1,0, 0)$ direction in the $(x,z)$-plane and the order parameter is calculated self-consistently [Fig.~\ref{flatbandszxplane} (a), (b)]. On the other hand, if the self-consistent changes in the order parameter are neglected, the $z$-component of the Zeeman field breaks the chiral symmetry $S_1$ and therefore the flat bands completely disappear [Fig.~\ref{flatbandszxplane} (c), (d)].

The existence of flat bands can  be easily generalized to  cases where the translationally invariant direction is not along $x$- or $y$-direction by projecting the nodes inside the Brillouin zone onto a line directed along the translationally invariant direction. Namely, the flat bands always appear between two nodes and the topological invariant remains the same 
as long as the effective one-dimensional Hamiltonians can be continuously deformed into each other without closing a gap.

\section{Topological phases of the Kitaev-Heisenberg model}

Next, we consider the situation where the triplet order parameter is a linear combination of three independent basis vectors.  To elucidate such kind of superconductor, we consider  the so-called Kitaev-Heisenberg model, which can be considered as a paradigmatic model that describes a large number of topological phases. 

We consider a honeycomb lattice with three inequivalent nearest neighbor bonds referred to as $\gamma = x,y,z$ [\onlinecite{comment}]. The undoped Kitaev-Heisenberg model in a Zeeman field is described by the Hamiltonian   
\begin{equation}
H = -J_K \sum_{\langle ij\rangle} S_i^\gamma S_j^\gamma + J_H \sum_{\langle ij\rangle} \vec{S}_i\cdot \vec{S}_j + \sum_i \mathbf{b} \cdot \vec{S}_i \ \ .
\label{hkh}
\end{equation}
The first term in the Hamiltonian (\ref{hkh}) is called  Kitaev interaction
\cite{Kitaev06}, and it describes an Ising-like  coupling between the $\gamma$
components of spins  $S_i^\gamma = \frac{1}{2} f^\dag_{i,
 \alpha}\sigma^\gamma_{\alpha\beta}f_{i,\beta}$  at each bond in the 
$\gamma$-direction. The second term describes an isotropic Heisenberg interaction
with interaction strength $J_H$. Because at half-filling the  system can be viewed as a Mott insulator, the doping effects can be taken into account by introducing a kinetic term
\begin{equation}
H_T=-t \sum_{\langle i j\rangle}\sum_{\sigma} f_{i, \sigma}^\dagger f_{j, \sigma} +h.c. 
\end{equation}
with the double occupancy prohibited similarly as in the $t$-$J$ model for high-$T_c$ cuprates. This constraint on the absence of double occupancies is of little importance in the limit of large doping. However, at small doping the constraint leads to an important  renormalization of the hopping amplitude, which is included in the  renormalized  amplitude  $t$ used in the following discussion. Thus,  the value of $t$ depends on the doping level.

The Kitaev-Heisenberg model displays a variety of interesting topological phases, which are summarized in Table \ref{tab:top_phases}. The topological spin liquid phases are the ones discovered by Kitaev \cite{Kitaev06}. Namely, in the absence of the Zeeman field the Kitaev model is equivalent to a tight-binding model for Majorana modes on a honeycomb lattice, and therefore exhibits topological flat bands similarly as graphene.  On the other hand, at finite Zeeman field the Kitaev spin liquid acquires a gap and has linearly dispersing Majorana edge modes and unpaired Majoranas at the vortices.

Already in the absence of a Zeeman field, the phase-diagram of the doped Kitaev-Heisenberg model  contains exotic superconducting phases. For dominating antiferromagnetic Heisenberg interaction $J_H>0$,  one obtains a topologically non-trivial time-reversal broken $d+id$ superconducting phase \cite{BlSc07} belonging to class C in the classification table for topological insulators and superconductors \cite{Schny+08}. On the other hand, a dominating ferromagnetic Kitaev interaction $J_K>0$ at small doping results in a time-reversal broken topologically nontrivial $p$-wave superconducting phase \cite{Vishwanath}. At intermediate and large doping the time-reversal symmetry is restored, and at large doping one finds a topologically nontrivial $\mathbb{Z}_2$ superconducting phase, which has one pair of counterpropagating edge modes \cite{Hyart12}.  On the other hand, at intermediate doping the superconductor is trivial in the $\mathbb{Z}_2$ classification \cite{Hyart12}. Nevertheless, an even number of symmetry protected edge states appear. The global phase-diagram  for the Kitaev-Heisenberg model (including different signs for $J_H$ and $J_K$) was recently computed in mean field theory by Okamoto \cite{Okamoto13}, and these results are supported by functional renormalization group calculations \cite{Scherer}.

%\begin{widetext}

%\end{widetext}

As we show below, in the presence of a Zeeman field, the time-reversal invariant $p$-wave superconducting phase can be tuned into i) a time-reversal broken topologically nontrivial phase supporting chiral Majorana edge modes and unpaired Majorana zero modes in the vortices, or ii) into a nodal $p$-wave superconductor supporting topological flat bands on its edges. We discuss these phases in the general framework of our Ginzburg-Landau theory, but we have also numerically checked all  results with the help of self-consistent mean field calculations for the Kitaev-Heisenberg model. These new topologically nontrivial superconducting phases are also included in Table \ref{tab:top_phases}.

\section{The various superconducting phases of the Kitaev-Heisenberg model in a Zeeman field}

We consider the Kitaev-Heisenberg model in the limit of intermediate and large doping, with dominating ferromagnetic Kitaev interaction. In this case, the linearized gap equations have three degenerate solutions \cite{Hyart12}. By projecting the Hamiltonian to a single band with dispersion $\xi_{1,(2)}=\pm |t(\mathbf{k})|-\mu$, these can be written as 
\begin{eqnarray}
\mathbf{d}_1(\mathbf{k})&=& \big[ \sin\big(\mathbf{\boldsymbol\delta}_2 \cdot \mathbf{k}-\varphi \big)- \sin\big(\mathbf{\boldsymbol \delta}_3 \cdot \mathbf{k}-\varphi\big) \big] \hat{\mathbf{e}}_x \nonumber\\
%%%%%%
\mathbf{d}_2(\mathbf{k})&=&\big[\sin\big(\mathbf{\boldsymbol \delta}_3 \cdot \mathbf{k}-\varphi\big) -  \sin\big(\mathbf{\boldsymbol \delta}_1 \cdot \mathbf{k}-\varphi \big)\big] \hat{\mathbf{e}}_y  \nonumber\\
\mathbf{d}_3(\mathbf{k})&=& \big[\sin\big(\mathbf{\boldsymbol \delta}_1 \cdot \mathbf{k}-\varphi \big)- \sin\big(\mathbf{\boldsymbol \delta}_2 \cdot \mathbf{k}-\varphi \big) \big] \hat{\mathbf{e}}_z. \label{basis}
\end{eqnarray}
Here, $t(\mathbf{k})=t\sum_j e^{i \mathbf{\boldsymbol \delta}_j \cdot \mathbf{k}}$, $\mathbf{\boldsymbol \delta}_1=(1/2, 1/2\sqrt{3})$,  $\mathbf{\boldsymbol \delta}_2=(-1/2, 1/2\sqrt{3})$ and $\mathbf{\boldsymbol \delta}_3=(0, -1/\sqrt{3})$ are the nearest neighbor vectors along the different links  and $\varphi=\arg[t(\mathbf{k})]$.

The Ginzburg-Landau theory in the presence of a Zeeman field is described by Eq.~(\ref{Free-energy}), where all coefficients are now non-zero. The parameters $\kappa_A$ and $\kappa_B$ control the magnitudes and phase-differences of the $\eta_i$. Since $\kappa_B >0$, the lowest energy solution in the absence of a Zeeman field is four-fold degenerate  $(\eta_1, \eta_2, \eta_3)=\eta_0 (1, \pm 1, \pm 1)$, see Ref.~\onlinecite{Hyart12}.  These solutions correspond to  $p+ ip$ and $p-ip$ order parameters for the different spin species, with the spin of the Cooper pair locked in the $(1,\pm1,\pm1)$ directions. The effect of a Zeeman field on the superconducting order parameter is described by the parameters $D_1$, $D_2^A$ and $D_2^B$.

The parameter $D_1$ is typically very small, because the contributions with $\pm \xi$ in expression (\ref{coeff}) approximately cancel each other around the Fermi surface. Therefore, we start by neglecting this term, and we only consider its effect in  special cases where it becomes important for our discussion (see below). In the absence of $D_1$, the analysis of the Ginzburg-Landau theory simplifies considerably, because the free-energy is always  minimized by $\eta_1, \eta_2, \eta_3 \in \mathbb{R}$. By further restricting ourselves to a Zeeman field with all $b_i \geq 0$, and taking into account that $D_2^B >0$, we notice that $\eta_1, \eta_2, \eta_3 \geq 0$.  Moreover, by using the  numerical observations that $\kappa_A \approx \kappa_B$ and $D_2^A \approx 2 D_2^B$, and by defining $\eta_0^2=-[t_I+t(|\mathbf{b}|)]/4 \kappa_A$, we can rewrite the free-energy as
\begin{eqnarray}
\frac{F}{N}&=& -[t_I+t(|\mathbf{b}|)] \eta_0^2  \bigg[ \sum_i \Big(-1+ \frac{D_2^A}{4 \kappa_A} \frac{b_i^2}{\eta_0^2} \Big)\frac{\eta_i^2}{\eta_0^2}  \nonumber\\&&  - \frac{D_2^A}{4 \kappa_A} \sum_{i<j} \frac{b_i}{\eta_0} \frac{b_j}{\eta_0} \frac{\eta_i}{\eta_0} \frac{\eta_j}{\eta_0}  + \frac{1}{4} \sum_i \frac{\eta_i^4}{\eta_0^4}  + \frac{1}{4} \sum_{i < j} \frac{\eta_i^2}{\eta_0^2} \frac{\eta_j^2}{\eta_0^2} \bigg]. \nonumber \\ \label{Free-energy-KH-Simp}
\end{eqnarray}
By straightforward calculation one can confirm that in the absence of Zeeman field, the minimization of Eq.~(\ref{Free-energy-KH-Simp}) gives an order parameter $(\eta_1, \eta_2, \eta_3)=\eta_0 (1, 1, 1)$.

Based on the expression (\ref{Free-energy-KH-Simp}), we expect that a Zeeman field with three nonzero components $b_i$ of comparable magnitude will choose an optimal direction in $(\eta_1, \eta_2, \eta_3)$-space, so that all the components $\eta_i$ will remain nonzero  until the Zeeman field reaches a critical value, where the superconductivity vanishes.  Moreover, the superconductor is generically expected to be fully gapped. These expectations are supported by our numerical calculations. The only exceptions are the specific values of the Zeeman field where topological phase transitions take place due to a change in the Chern number (see below). Similarly, as in the case of the helical $p$-wave superconductor with two basis vectors consireded in section III, we find that the self-consistent changes in the order parameter are not important in the description of these topological phase transitions.

On the other hand, we may expect the self-consistent changes to become important if only two components of the Zeeman field are nonzero. To understand the effect of a Zeeman field on the superconducting order parameter in these cases, we consider a specific Zeeman field $\mathbf{b}=b(1,0,1)/\sqrt{2}$. In this case, it is useful to make a transformation $\eta_1=(\eta_++\eta_-)/\sqrt{2}$ and $\eta_3=(\eta_++\eta_-)/\sqrt{2}$, which diagonalizes the quadratic part of the free-energy Eq.~(\ref{Free-energy-KH-Simp}). After this transformation, one notices that both $\eta_+$ and $\eta_-$ have larger masses than $\eta_2$, so that one can expect them to vanish before $\eta_2$ at sufficiently large Zeeman fields. By minimizing the resulting free-energy, we find that for $b<\eta_0 \sqrt{8\kappa_A/D_2^A}$
\begin{eqnarray}
\eta_1&=&\eta_3=\eta_0 \sqrt{1-\frac{D_2^A}{8\kappa_A} \frac{b^2}{\eta_0^2}}, \nonumber \\
\eta_2&=&\eta_0 \sqrt{1+\frac{D_2^A}{8\kappa_A} \frac{b^2}{\eta_0^2}}
\end{eqnarray}
and for $b>\eta_0 \sqrt{8\kappa_A/D_2^A}$
\begin{eqnarray}
\eta_1&=&\eta_3=0, \nonumber \\
\eta_2&=&\eta_0 \sqrt{2}.
\end{eqnarray}
Once $\eta_1=\eta_3=0$, the chiral symmetry $S_2$ defined in Eq.~(\ref{S2}) is automatically satisfied. In addition, due to the $p$-wave nature of the basis functions, the superconductor becomes  gapless at this value of the Zeeman field, and we therefore  can identify $b_{c, \rm{ch}}=\eta_0 \sqrt{8\kappa_A/D_2^A}$. 

If the Zeeman field is rotated away from the $(1, 0, 1)$-direction but stays in the $(x,z)$-plane, the masses of the eigenmodes which diagonalize the quadratic part of the free-energy become more asymmetric. In particular, when one approaches the $(0,0,1)$ or $(1,0,0)$ direction the mass of one of these eigenmodes approaches the mass of $\eta_2$, and therefore we can expect that $b_{c, \rm{ch}}$ is smallest at $(1,0,1)$-direction and becomes larger when approaching $(0,0,1)$ and $(1,0,0)$-directions. By minimizing the free-energy (\ref{Free-energy-KH-Simp}) In the extreme limit of $\mathbf{b}=b (0,0,1)$, we find that for $b<\eta_0 \sqrt{4\kappa_A/3 D_2^A} $
\begin{eqnarray} 
\eta_1&=&\eta_2=\eta_0  \sqrt{1+\frac{D_2^A}{4\kappa_A} \frac{b^2}{\eta_0^2}}, \nonumber \\
\eta_3&=&\eta_0  \sqrt{1-\frac{3D_2^A}{4\kappa_A} \frac{b^2}{\eta_0^2}}
\end{eqnarray}
and for $b>\eta_0 \sqrt{4\kappa_A/3 D_2^A}$
\begin{eqnarray} 
\eta_1&=&\eta_2=\eta_0  \sqrt{4/3}, \nonumber \\
\eta_3&=&0.
\end{eqnarray}
Importantly, the chiral symmetry $S_2$ is no longer necessarily satisfied even for $b>\eta_0 \sqrt{4\kappa_A/3 D_2^A}$, because two of the components of the order parameter are now non-zero. However, it is possible to show that the chiral symmetry exists if the phase-difference of $\eta_1$ and $\eta_2$ is fixed to a particular value by the previously neglected $D_1$-term. To understand in detail how the phase-difference depends on the strength of the Zeeman field, we now assume that $\eta_1=\eta_0  \sqrt{4/3} e^{i\phi_1}$, $\eta_2=\eta_0  \sqrt{4/3} e^{i\phi_2}$, $\eta_3=0$ and minimize the free-energy Eq.~(\ref{Free-energy}) with respect to phase-difference $\phi=\phi_1-\phi_2$. The phase-dependent terms of the free-energy can in this case be written as
\begin{equation}
\frac{F_\phi}{N}=
-\bigg(\frac{4}{3}\bigg)^2\kappa_A \eta_0^4  \cos(2\phi)
+ \frac{8}{3}  D_1  \eta_0^2 b  \sin(\phi).   
\end{equation}
For small values of $b$ this free-energy would be minimized by $\phi=0$, and by increasing $b$, the phase difference $\phi$ moves towards $\pm \pi/2$, where the sign is determined by the sign of $D_1$. For Zeeman fields $b  >8 \kappa_A \eta_0^2/3|D_1|$, the phase-difference becomes pinned to $\pm \pi/2$. In this case, the superconductor becomes nodal and the chiral symmetry $S_2$ is satisfied. Because $\eta_0$ decreases with increasing $|\mathbf{b}|$, this inequality will eventually be reached, but at this point the magnitude of the superconducting order parameter is already very small, because typically $\kappa_A \eta_0 (\mathbf{b}=0)/|D_1| \gg 1$. Therefore, although the chiral symmetry always emerges for all in-plane Zeeman fields, the critical field $b_{c, \rm{ch}}$ depends strongly on the direction.

\subsection{Time-reversal broken topological superconductivity}

\begin{figure}[tbp]
\centering\includegraphics[width=8.5cm]{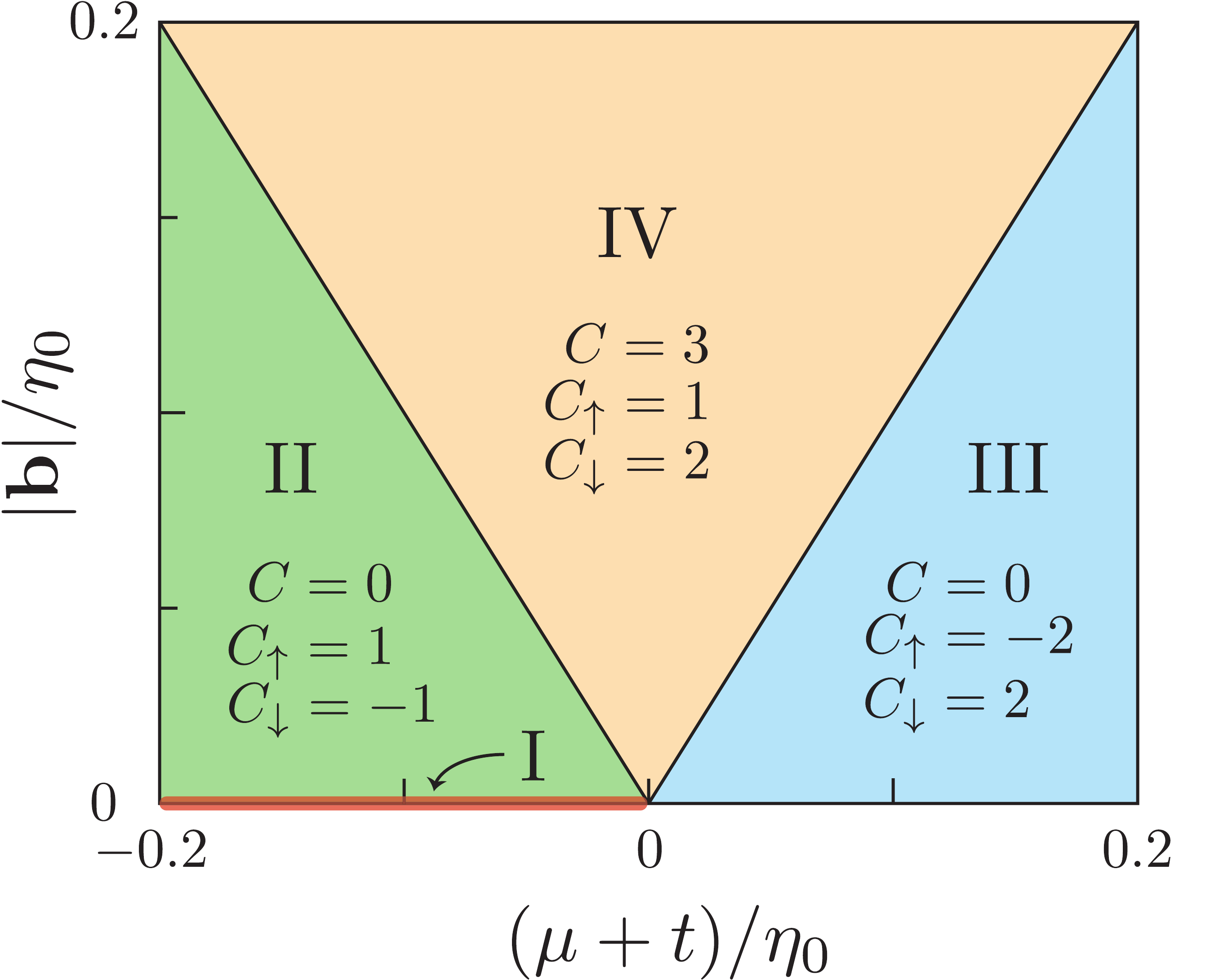}
\caption{Phase diagram of the doped Kitaev model near  $\mu= - t$ as a function of Zeeman field oriented in the $(1,1,1)$ direction. The four phases are I: time reversal invariant topologically nontrivial $\mathbb{Z}_2$ superconducting phase \cite{Hyart12}, II$\&$III: time reversal broken $p$-wave superconductor with symmetry protected edge modes, IV: time reversal broken topologically non-trivial $p$-wave superconducting phase. In phases II and III, a Zeeman field rotated away from the $[1,\pm 1, \pm1]$ directions will break the symmetry protection and hybridise the edge modes. 
}
\label{phasefig}
\end{figure}

%Description of the phases

In order to discuss the different fully gapped topological phases of the model, we first note that if the Zeeman field points in $(1, s_2, s_3)$ directions with $s_2, s_3=\pm1$,  the free energy (\ref{Free-energy-KH-Simp}) is minimized by the order parameter $(\eta_1, \eta_2, \eta_3)=\eta_0 (1, s_2, s_3)$, i.e.~each direction of the Zeeman field favors one of the degenerate solutions found in the absence of the Zeeman field. As discussed in Ref.~[\onlinecite{Hyart12}], a global spin rotation allows a transformation of the Hamiltonian into a form where  $\mathbf{d}$ remains in ($x,y$)-plane. Moreover, in the transformed frame, the Zeeman field  points in the $z$-direction, so that  the fermions with different spins are decoupled. The block diagonal form of the Hamiltonian, obtained this way for the specific directions of the Zeeman field, is useful for  understanding  the different topological phases, similarly as in the case of the helical $p$-wave superconductor with two basis vectors discussed in section III.

In the absence of the Zeeman field there are two topologically distinct phases, due to a change of the Fermi surface topology with doping \cite{Hyart12}. The topologically nontrivial $\mathbb{Z}_2$ superconducting phase exists for $|\mu|>t$ (the horizontal line I in Fig.~\ref{phasefig}), whereas the superconductor is topologically trivial in the $\mathbb{Z}_2$ classification for $|\mu|<t$. Nevertheless, the Chern numbers $C_\uparrow$ and $C_{\downarrow}$ are nonzero also in the topologically trivial phase, and therefore an even number of edge states exists as long as the two spin blocks are decoupled. The protection of the edge states in the trivial phase is not guaranteed by  time-reversal symmetry alone, but requires the existence of the microscopic symmetry which allowed to block diagonalize the Hamiltonian.  

Once the Zeeman field is turned on, the time-reversal symmetry is broken and the $\mathbb{Z}_2$ classification does not exist anymore. However, as long as the microscopic symmetry is valid, $C_\uparrow$ and $C_{\downarrow}$ are well-defined and allow to determine the number of the edge states. Moreover, the energy gap can close at the high-symmetry points of the Brillouin zone, allowing the Chern number $C=C_\uparrow+C_\downarrow$ to become nonzero. Similar to the case of section III,  we concentrate on the $M$ points, because there the gap closings happen at reasonable carrier densities so that the superconducting order parameter can be large.  Due to the symmetries of the honeycomb lattice, the gap closings happen simultaneously in all three $M$ points $M_1=(0, 2\pi/\sqrt{3})$, $M_2=(\pi, \pi/\sqrt{3})$ and $M_3=(\pi, -\pi/\sqrt{3})$. Because $|\mathbf{d}(M_i)|=0$ is automatically satisfied at these points, and since $|t(M_i)|=t$, the gap closings take place at chemical potentials $\mu=\pm t \pm |\mathbf{b}|$. We now concentrate on the topological phase transitions taking place in the vicinity of $\mu \approx -t$.  We find that $C_{\uparrow}=1$ for $\mu<-t +|\mathbf{b}|$ and $C_{\uparrow}=-2$ for $\mu>-t+|\mathbf{b}|$, whereas $C_{\downarrow}=-1$  for $\mu<-t-|\mathbf{b}|$ and $C_{\downarrow}=2$ for $\mu>-t-|\mathbf{b}|$. This way, we arrive at the phase-diagram shown in Fig.~\ref{phasefig}. One important difference to the earlier case is that now the time-reversal broken topologically nontrivial phase with $C=3$ (region IV in Fig.~\ref{phasefig}) supports unpaired Majoranas in the vortices and an odd number of chiral Majorana edge modes. Although in \fig{phasefig} we show the parameter space for the time-reversal broken topologically nontrivial phase only for the $(1,1,1)$ direction, we find that this phase exists for a wide range of directions and magnitudes of the Zeeman field.

\subsection{Majorana flat bands} 

\begin{figure}
\includegraphics[scale=0.35]{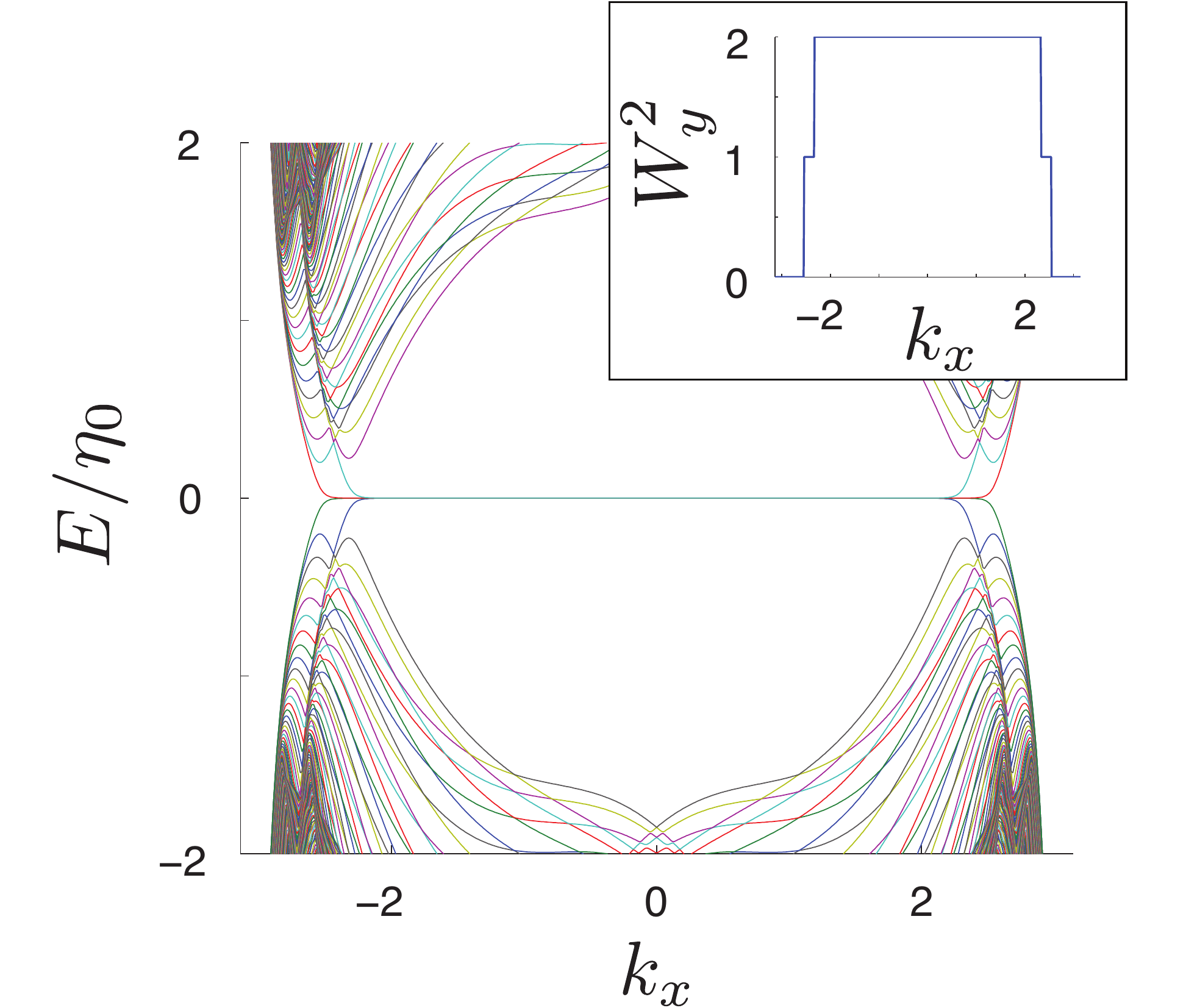}
\caption{Kitaev-Heisenberg model: Majorana flat bands and the corresponding topological invariant for edges along the x-direction. The parameters were chosen as $\mu = -1.4 t$, $b_1=b_3=\sqrt{2} \eta_0$, $b_2=0$. For these parameters, the self-consistently calculated order parameter is $\eta_1=\eta_3=0$ and $\eta_2=\sqrt{2} \eta_0$. The flatbands appear when the order parameter is computed self-consistently in the presence of a Zeeman field, and are absent otherwise. 
  }
\label{flatbands}
\end{figure}

We now concentrate on the nodal superconducting phases with an emergent chiral symmetry that can appear in the case of in-plane Zeeman fields.  As discussed above,  the magnitude of the critical field $b_{c, \rm{ch}}$ depends on the direction of the Zeeman field, and the smallest critical field $b_{c, \rm{ch}}=\eta_0 \sqrt{8\kappa_A/D_2^A}$ is obtained when the Zeeman field points along the $(1,0,1)$ direction. Therefore, we now concentrate on this case. By using Eqs.~(\ref{coeff}), we estimate that $D_2^A/\kappa_A \approx 2$ and additionally we set $\eta_0=0.05 t$ in the following calculations. This way we obtain the Majorana flat bands shown in Fig.~\ref{flatbands}. We emphasize that these flat bands are absent if the order parameter is not calculated self-consistently.

%Summary and discussion

\section{Conclusions and discussion}

In summary, we have have developed an intuitive and reasonably general Ginzburg-Landau theory for description of the self-consistent changes of the superconducting order parameter in the presence of the Zeeman field, and we have studied the effect of a Zeeman field on fully gapped time-reversal invariant superconducting phases. By calculating the phase diagrams as a function of the strength and direction of the Zeeman field, we have shown that the time-reversal invariant helical $p$-wave superconducting phase can be tuned into a time reversal broken topologically nontrivial phase supporting chiral edge states, or a nodal $p$-wave superconductor supporting topological flat bands on its edges. 

In this work, we have assumed that the superconducting order parameter is spatially homogeneous. In the absence of a Zeeman field this assumption has been confirmed with a self-consistent mean field theory for the Kitaev-Heisenberg model \cite{Kimme}. However, in the presence of a Zeeman field, interesting effects such as FFLO phases \cite{Fulde64, Larkin65} can in principle appear. Additionally we have not considered the effect of Rashba spin-orbit coupling and the possibility of a mixture of singlet and triplet superconducting order parameters, which can appear in noncentrosymmetric superconductors \cite{Balatsky}, and may result in breaking of the chiral symmetry and appearance of unidirectional edge modes \cite{Lee12}. We also want to point out, that our general Ginzburg-Landau theory may be utilized also in completely different research directions such as in the identification of the order parameter of unconventional superconductors, which  go beyond the scope of the present work.

\acknowledgements

T.H. thanks the  Dutch Science Foundation NWO/FOM, and B.R. acknowledges support by DFG.


\begin{thebibliography}{99}

%Topological materials
\bibitem{Thouless}
D. J. Thouless et al., Phys. Rev. Lett. {\bf 49}, 405 (1982).
\bibitem{Wen}
X.G. Wen, Quantum Field Theory of Many-Body Systems: From the Origin of Sound to an Origin of Light and Electrons, (Oxford University Press, 2007).
\bibitem{Volovik-book}
G. E. Volovik, Universe in a Helium Droplet, (Oxford University Press, 2003).
\bibitem{HaKa10} M.Z.~Hasan and C.L.~Kane, Rev. Mod. Phys. {\bf 82}, 3045 (2010).
\bibitem{Zhang-review}
X.-L. Qi and S.-C. Zhang, Rev. Mod. Phys. {\bf 83}, 1057 (2011).


%Kitaev spin liquid
\bibitem{Kitaev06} A. Kitaev, Ann. Phys. \textbf{321}, 2 (2006).


%Classification 
\bibitem{Schny+08} A.P. Schnyder, S. Ryu, A. Furusaki, and A. W. Ludwig, 
Phys. Rev. B \textbf{78}, 195125 (2008). 
\bibitem{Kitaev-class} 
A. Yu. Kitaev, AIP Conf. Proc. {\bf 1134}, 22 (2009).


%Majorana reviews
\bibitem{Lejinse12}
M. Leijnse and K. Flensberg, Semicond. Sci. Technol. {\bf 27}, 124003 (2012).
\bibitem{Alicearev}
J. Alicea, Rep. Prog. Phys. {\bf 75}, 076501 (2012).
\bibitem{Beenakker13} 
C.W.J. Beenakker, Annu. Rev. Con. Mat. Phys. {\bf 4}, 113 (2013).


%Non-Abelian statistics
\bibitem{MooreRead}
G. Moore and N. Read, Nucl. Phys. B {\bf 360}, 362 (1991).
\bibitem{ReadGreen}
N. Read and D. Green, Phys. Rev. B {\bf 61}, 10267 (2000).
\bibitem{Ivanov}
D. A. Ivanov, Phys. Rev. Lett. {\bf 86}, 268 (2001).

%p+ip
\bibitem{KopninSalomaa91}
N. B. Kopnin and M. M. Salomaa, Phys. Rev. B {\bf 44}, 9667 (1991).
\bibitem{Volovik}
G. Volovik, JETP Letters {\bf 70}, 609 (1999).
\bibitem{Wimmer}
M. Wimmer, A. R. Akhmerov, M. V. Medvedyeva, J. Tworzydlo, and C. W. J. Beenakker, Phys. Rev. Lett. {\bf 105}, 046803 (2010).

%spinful p+ip and Zeeman field
\bibitem{DasSarma06}
S. Das Sarma, C. Nayak, and S. Tewari, Phys. Rev. B {\bf 73}, 220502(R) (2006).
\bibitem{Lee09}
P. A. Lee, arXiv:0907.2681 (2009).

%Surface of TI
\bibitem{FuKane08}
L. Fu and C. L. Kane,  Phys. Rev. Lett. {\bf 100}, 096407 (2008).

%two-dimensional systems (s-wave+ spin orbit + Zeeman field)
\bibitem{Sato09SO}
M. Sato, Y. Takahashi, and S. Fujimoto,  Phys. Rev. Lett. {\bf 103}, 020401 (2009).
\bibitem{Sau10}
J. D. Sau, R. M. Lutchyn, S. Tewari, and S. Das Sarma, Phys. Rev.
Lett. {\bf 104}, 040502 (2010).

%one-dimensional systems
\bibitem{Kitaev-wire}
A. Yu. Kitaev, Phys.-Usp. {\bf 44}, 131 (2001).
\bibitem{FuKane09}
L. Fu and C. L. Kane,  Phys. Rev. B {\bf 79}, 161408(R) (2009).
\bibitem{Lutchyn10}
R. M. Lutchyn, J. D. Sau, and S. Das Sarma, Phys. Rev. Lett. {\bf 105}, 077001 (2010).
\bibitem{Oreg}
Y. Oreg, G. Refael, and F. von Oppen,  Phys. Rev. Lett. {\bf 105}, 177002 (2010).

%Superconducting phases in Kitaev Heisenberg model
\bibitem{BuNa11}  
F.J.~Burnell and C.~Nayak,  Phys. Rev. B {\bf 84}, 125125 (2011).
\bibitem{Hyart12}
T. Hyart, A.R. Wright, G. Khaliullin, and B. Rosenow, Phys. Rev. B, \textbf{85}, 140510(R) (2012).
\bibitem{Vishwanath}
Y. You, I. Kimchi, and A. Viswanath, Phys. Rev. B {\bf 86}, 085145 (2012).
\bibitem{Okamoto13}
S. Okamoto, Phys. Rev. B {\bf 87}, 064508 (2013).
\bibitem{Scherer}
D. D. Scherer, M. M. Scherer, G. Khaliullin, C. Honerkamp, B. Rosenow, arXiv:1403.6762.

%TRS invariant superconductors (p+ip, p-ip)
\bibitem{Roy08}
R. Roy, arXiv:0803.2868.
\bibitem{Qi2009}
X.-L. Qi, T. L. Hughes, S. Raghu, and S.-C. Zhang, Phys Rev. Lett. {\bf 102}, 187001 (2009).
\bibitem{Sato09a}
M. Sato, Phys. Rev. B {\bf 79}, 214526 (2009).
%%%%%%%%%%%%%%%%%%%%%%%%%%%%%%%%%%
\bibitem{Sato09} 
M. Sato and S. Fujimoto, Phys. Rev. B {\bf 79}, 094504 (2009).
%Central paper. Explains how magnetic field can cause phase-transition from TRS invariant SC with topologically nontrivial phase to a TRS-broken topologically nontrivial SC in noncentrosymmetric superconductors. Also symmetry protected edge states originating from Z2 phase in the presence of magnetic field.
%%%%%%%%%%%%%%%%%%%%%%%%%%%%%%%%%%
\bibitem{Qi2010}
X.-L. Qi, T.L. Hughes, and S.-C. Zhang,  Phys. Rev. B {\bf 81}, 134508 (2010).
\bibitem{Sato10} M. Sato, Phys. Rev. B \textbf{81}, 220504(R) (2010).
\bibitem{YTada}
Y. Tada, N. Kawakami and S. Fujimoto, New J. Phys. {\bf 11}, 055070  (2009).
\bibitem{Chen14}
X. Chen, Y. Yao, H. Yao, F. Yang, J. Ni, arXiv:1404.3346.

%quantum computation
\bibitem{Nayak+08} C.~Nayak, S.H.~Simon, A.~Stern, M.~Freedman, and S.~Das Sarma, Rev. Mod. Phys. {\bf 80}, 1083 (2008).
\bibitem{Alicea}
J. Alicea, Y. Oreg, G. Refael, F. von Oppen, and M. P. A. Fisher, Nat. Phys. {\bf 7}, 412 (2011).
\bibitem{Hassler11}
F. Hassler, A. R. Akhmerov, and C. W. J. Beenakker, New. J. Phys. {\bf 13}, 095004 (2011).
\bibitem{Hyart13}
T. Hyart, B. van Heck, I. C. Fulga, M. Burrello, A. R. Akhmerov, C. W. J. Beenakker, 	Phys. Rev. B {\bf 88}, 035121 (2013).


%Superfluid Helium
\bibitem{VollhardtWolfle}
D. Vollhardt and P. W\"olfle, {\em The  Superfluid Phases of Helium 3},  Taylor \& Francis  (1990).


%Kitaev-Heisenberg model
\bibitem{JaKh09} G. Jackeli and G. Khaliullin,
Phys. Rev. Lett. {\bf 102}, 017205 (2009).
\bibitem{Cha+10} J.~Chaloupka, G.~Jackeli, and G.~Khaliullin, Phys. Rev. Lett. {\bf 105}, 027204 (2010).



\bibitem{Silaev}
See Yu. Makhlin, M. Silaev, G.E. Volovik, 	arXiv:1312.2677, and references therein.


%Review of superconductivity and GL theory
\bibitem{Sigrist}
M. Sigrist and K. Ueda, Rev. Mod. Phys. {\bf 63}, 239 (1991).


\bibitem{Mannhart}
L. Li,	 C. Richter, J. Mannhart and R. C. Ashoori, Nat. Phys. {\bf 7}, 762 (2011).

\bibitem{Moler}
J. A. Bert,	 B. Kalisky, C. Bell,  M. Kim, Y. Hikita, H. Y. Hwang and K. A. Moler, Nat. Phys. {\bf 7}, 767 (2011).

\bibitem{Pickett}
W. E. Pickett, R. Weht, and A. B. Shick,  Phys. Rev. Lett. {\bf 83}, 3713 (1999).

\bibitem{Saxena}
S. S. Saxena \textit{et al.}, Nature {\bf 406}, 587 (2000).



%Gappless topological superconductors


%Magnetic field induced flat bands
\bibitem{Lee12} 
C. L. M. Wong, J. Liu, K. T. Law and P. A. Lee, Phys. Rev. B {\bf 88}, 060504(R) (2013).



%Intrinsic gappless topological superconductors and semimetals

%review
\bibitem{Heikkila-flat-bands} 
T. T. Heikkil\"{a}, N. B. Kopnin, G. E. Volovik, JETP Lett. {\bf 94}, 233 (2011).

%Flat bands graphene
\bibitem{Nakada96} K. Nakada et al., Phys. Rev. B {\bf 54}, 17954 (1996).
\bibitem{Fujita96} M. Fujita et al., J. Phys. Soc. Jpn. {\bf 65}, 1920 (1996).

%Flat bands in d_{x^2-y^2}
\bibitem{CuTheory} C.-R. Hu, Phys. Rev. Lett. \textbf{72}, 1526 (1994).

%Early discussion of topology in context of flat bands
\bibitem{Ryu02}
S. Ryu and Y. Hatsugai, Phys. Rev. Lett. {\bf 89}, 077002 (2002).

%Flat bands in vortices
\bibitem{Volovik-vortices}
G. E. Volovik, JETP Lett. {\bf 93}, 66 (2011).

%Flat bands noncentrosymmetric
\bibitem{Tanaka10}
Y. Tanaka, Y. Mizuno, T. Yokoyama, K. Yada, M. Sato Phys. Rev. Lett. {\bf 105}, 097002 (2010).
\bibitem{SchnyderRyu11} 
A. P. Schnyder and S. Ryu, Phys. Rev. B {\bf 84}, 060504 (2011). 
\bibitem{Brydon11}
P. M. R. Brydon, A. P. Schnyder, and C. Timm, 
Phys. Rev. B {\bf 84}, 020501 (2011). 
\bibitem{Sato11}
M. Sato, Y. Tanaka, K. Yada and T. Yokoyama, Phys. Rev. B {\bf 83}, 224511 (2011). 
\bibitem{Schnyder12}
A. P. Schnyder, P. M. R. Brydon, and C. Timm, 
Phys. Rev. B {\bf 85}, 024522 (2012).

\bibitem{comment2}
The $Z_2$ invariant can be defined even if the Hamiltonian is not block diagonal, but the two definitions are equivalent in the present case.

\bibitem{Manmana12}
S. R. Manmana, A. M. Essin, R. M. Noack and V. Gurarie,  Phys. Rev. B {\bf 86}, 205119 (2012).

\bibitem{Tewari}
S. Tewari and J. D. Sau, Phys. Rev. Lett. {\bf 109}, 150408 (2012).

\bibitem{comment}
We use the conventions defined in Fig.1 in Ref.~\onlinecite{Hyart12}.

\bibitem{BlSc07} A.M. Black-Schaffer and S. Doniach, 
Phys. Rev. B \textbf{75}, 134512 (2007). 




\bibitem{Kimme}
L. Kimme, M.Sc. Thesis, University of Leipzig (2012).

\bibitem{Fulde64}
P. Fulde and R. A. Ferrell, Phys. Rev. {\bf 135}, A550 (1964). 

\bibitem{Larkin65}
A. I. Larkin and Yu. N. Ovchinnikov, Sov. Phys. JETP {\bf 20}, 762 (1965).

\bibitem{Balatsky}
Y. Tanaka, T. Yokoyama, A. V. Balatsky, and N. Nagaosa, Phys. Rev. B {\bf 79}, 060505(R) (2009).

\end{thebibliography}
\end{document}